\documentclass[twocolumn,twocolappendix]{aastex63}

\usepackage{graphicx}
\usepackage{txfonts}
\usepackage{color}
\usepackage{afterpage}
\usepackage{ulem}
\usepackage{rotating}

\defcitealias{Currie2022}{C22}
\defcitealias{shi24}{SM24}

\shorttitle{Prediction of dust emission from the CPD of AB~Aurigae~b}
\shortauthors{Shibaike et al.}

\begin{document}

\title{Predictions of Dust Continuum Emission from a Potential Circumplanetary Disk: A Case Study of the Planet Candidate AB Aurigae b}

\correspondingauthor{Yuhito Shibaike}
\email{yuhito.shibaike@nao.ac.jp}

\author[0000-0003-2993-5312]{Yuhito Shibaike}
\affil{ALMA Project, National Astronomical Observatory of Japan, 2-21-1 Osawa, Mitaka, Tokyo 181-8588, Japan}
\affil{Space Research and Planetary Sciences, Physics Institute, University of Bern, CH-3012 Bern, Switzerland}
\author[0000-0002-3053-3575]{Jun Hashimoto}
\affil{Astrobiology Center, National Institutes of Natural Sciences, 2-21-1 Osawa, Mitaka, Tokyo 181-8588, Japan}
\affil{Subaru Telescope, National Astronomical Observatory of Japan, Mitaka, Tokyo 181-8588, Japan}
\affil{Department of Astronomy, School of Science, Graduate University for Advanced Studies (SOKENDAI), Mitaka, Tokyo 181-8588, Japan}
\author[0000-0001-9290-7846]{Ruobing Dong}
\affil{Department of Physics and Astronomy, University of Victoria, Victoria, BC V8P 5C2, Canada}
\affil{Kavli Institute for Astronomy and Astrophysics, Peking University, Beijing 100871, China}
\author[0000-0002-1013-2811]{Christoph Mordasini}
\affil{Space Research and Planetary Sciences, Physics Institute, University of Bern, CH-3012 Bern, Switzerland}
\author[0000-0003-1117-9213]{Misato Fukagawa}
\affil{ALMA Project, National Astronomical Observatory of Japan, 2-21-1 Osawa, Mitaka, Tokyo 181-8588, Japan}
\author{Takayuki Muto}
\affil{Division of Liberal Arts, Kogakuin University, 2665-1, Nakano-cho, Hachioji-chi, Tokyo, 192-0015, Japan}


\begin{abstract}
Gas accreting planets embedded in protoplanetary disks are expected to show dust thermal emission from their circumplanetary disks (CPDs). However, a recently reported gas accreting planet candidate, AB Aurigae b, has not been detected in (sub)millimeter continuum observations. We calculate the evolution of dust in the potential CPD of AB Aurigae b and predict its thermal emission at 1.3 mm wavelength as a case study, where the obtained features may also be applied to other gas accreting planets. We find that the expected flux density from the CPD is lower than the $3\sigma$ level of the previous continuum observation by ALMA with broad ranges of parameters, consistent with the non-detection. However, the expected planet mass and gas accretion rate are higher if the reduction of the observed near-infrared continuum and H$\alpha$ line emission due to the extinction by small grains is considered, resulting in higher flux density of the dust emission from the CPD at (sub)millimeter wavelength. We find that the corrected predictions of the dust emission are stronger than the $3\sigma$ level of the previous observation with the typical dust-to-gas mass ratio of the inflow to the CPD. This result suggests that the dust supply to the vicinity of AB Aurigae b is small if the planet candidate is not the scattered light of the star but is a planet and has a CPD. Future continuum observations at shorter wavelength are preferable to obtain more robust clues to the question whether the candidate is a planet or not.
\end{abstract}



\section{Introduction} \label{sec:intro}
When a gas giant planet forms in a protoplanetary disk (PPD), the embedded planet accretes gas from the disk. Understanding of the gas accretion process is essential for the planet formation theory, since gas giant planets dominate the formation of the whole system by their large mass. Historically, previous research has been working on the gas accretion from the theoretical aspects, using numerical hydrodynamical simulations and analytical interpretation. They have revealed that the planet--disk interactions lead to open a gap around the planetary orbit in the gas disk \citep[e.g.,][]{lin1986}, and inside the gap, the gas accreting planet forms a small gas disk around the planet, which is called as a circumplanetary disk (CPD), if the gas around the planet is cool enough \citep[e.g.,][]{lubow99}.

Recently, observations of gas accreting planets have also been possible due to the development of the ability of telescopes. There have been several reports of the detection of the spacial concentration of near-infrared continuum emission in multiple systems, PDS~70~b and c \citep{wagn18a,haff19a}, AB~Aurigae~b (AB~Aur~b) \citep[][hereafter \citetalias{Currie2022}]{Currie2022}, HD~169142~b \citep{ham23}, and NWC~758~c \citep{wag23}, which have been considered as gas accreting planets. Especially, H$\alpha$ line emission has also been detected from PDS~70~b and c, and AB~Aur~b, supporting that they are forming planets \citep{Zhou2021a}, because the detection of H$\alpha$ line emission can be interpreted that the planets accrete gas and make shocks heating up the surfaces of the planets (or the CPDs) enough to emit H$\alpha$ line emission.

One of the other ways to support the existence of a forming planet is to detect the dust continuum emission from the planet candidate at (sub)millimeter wavelength. This is because dust particles can grow large to millimeter size by their mutual collisions in CPDs. However, there has been only one clear detection of a CPD, the CPD of PDS~70~c, where its dust thermal emission has been observed by Atacama Large Millimeter/submillimeter Array (ALMA) at Band 7 ($\lambda=855~\mu{\rm m}$) \citep{isell19a,Benisty2021}. The previous observations have not detect the dust continuum from PDS~70~b at any (sub)millimeter. \citet{Tang2017} also observed the AB~Aur system by the continuum emission at Band 6 ($\lambda=1.3~{\rm mm}$), and the dust emission from AB~Aur~b was not detected either (see Appendix \ref{sec:previous}). There are also few candidates, but they have not obtained the consensus of the research field \citep{and21,bae22}.

A recent work, \citet[][hereafter \citetalias{shi24}]{shi24}, developed a model predicting the flux density of the thermal emission of dust in CPDs considering the in-situ evolution of the dust in the disks and investigated the reason why PDS~70~c has detection of dust continuum but PDS~70~b does not. They concluded that the difference of the amount of dust supply to the vicinity of the planets can be the reason, and it could also be the reason why the observed H$\alpha$ emission from PDS~70~b is stronger than that of PDS~70~c by considering the reduction of the observed H$\alpha$ emission by the extinction caused by small dust grains around the planets. On the other hand, the reason of the non-detection of the dust emission from the potential CPD of AB~Aur~b has not been investigated.

Therefore, in this work, we first show the updates of the evolution and emission model of dust in CPDs and then show the predictions of the dust evolution and emission of the potential CPD of AB~Aur~b as a case study. The trends and features of the predictions may also be applied to the potential CPDs of other gas accrediting planets. Studying the potential CPD of AB~Aur~b is important also from the point of view that there is still discussion about the origin of the near-infrared and H$\alpha$ emission of AB~Aur; the emission is either from a gas accreting planet or scattered light of the central star at the surface of the PPD \citep[e.g.,][]{zho22}.

In Section \ref{sec:methods}, we explain the updates of the model. We then show the gas and dust distribution in the CPD of AB~Aur~b calculated by our model in Section \ref{sec:distribution}. In Section \ref{sec:monomer}, we investigate the effects of the monomers' conditions, which make the dust particles more fragile. In Section \ref{sec:band6}, we predict the flux density of the dust thermal emission at Band 6 and compere the results with the previous observation. In Section \ref{sec:extincstion}, we also predict the flux density of the dust emission considering the upward revision of the expected planet mass and the gas accretion rate of AB~Aur~b by the reduction of the observed rear-infrared and H$\alpha$ emission due to the extinction by small grains. We then discuss the reasons of the non-detection of the CPD of AB~Aur~b based on our results in Section \ref{sec:planet}. We also show the prediction of the emission at ALMA Band 7 and propose future observations in Section \ref{sec:band7}. We conclude this work in Section \ref{sec:conclusions}.

\section{Methods} \label{sec:methods}
\citetalias{shi24} developed a model predicting the flux density of the dust thermal emission from a CPD of a gas accreting planet. Here, we explain the differences and updates from the original model to avoid duplication. See Section 2 of \citetalias{shi24} for the detailed explanations.

\subsection{Gas disk} \label{sec:gasdisk}
We consider a steady-state viscous accretion CPD with continuous gas infall from the parental PPD. The radial distribution of the gas surface density $\Sigma_{\rm g}$ and the midplane temperature $T$ is calculated by a model improved by the ``gas-staved disk'' model, originally proposed by \citet{can02}. The gas profiles mainly depend on three input parameters: the planet mass ($M_{\rm p}$), the gas accretion rate ($\dot{M}_{\rm g}$), and the strength of turbulence in the CPD ($\alpha$). In this work, we first use the values of the planet mass and the gas accretion rate estimated in \citetalias{Currie2022}. We then consider the reduction effects of the obtained values of the planet properties due to the extinction by small grains in Section \ref{sec:extincstion}.

One of the most important properties determining $F_{\rm d}$, the total flux density of the thermal emission of the dust in a CPD, is the surface area of the ``dust-containing region'' of the CPD as we will explain in the later sections. In the model, gas and dust flow onto the CPD at $r\leq r_{\rm inf}$ (we refer to the region as the ``inflow region''), and dust only moves inward by the radial drift. Therefore, the dust-containing region in this model is also expressed as $r\leq r_{\rm inf}$. The inflow region is determined by the angular momentum of the gas inflow. The outer edge of the inflow region is, $r_{\rm inf}=25/16~r_{\rm c}$, where $r_{\rm c}\equiv j_{\rm c}^{2}/(GM_{\rm p})$ is the centrifugal radius of the inflowing gas with the average specific angular momentum, $j_{\rm c}$ \citep{can02,war10}. We define $j_{\rm c}\equiv l\Omega_{\rm p}R_{\rm H}^{2}$, where $l$ is the angular momentum bias of the gas inflow, $\Omega_{\rm p}=\sqrt{GM_{\rm *}/a_{\rm p}^{3}}$ is the Keplerian frequency of the planet, and $R_{\rm H}\equiv(M_{\rm p}/(3M_{*}))^{1/3}a_{\rm p}$ is the Hill radius. The letter $G$ is the gravitational constant and $a_{\rm p}$ is the planetary orbit. We use an approximation to calculate $l$ obtained by a fitting to the results of the previous hydrodynamical simulations in \citet{war10}:
\begin{equation}
l=0.12\left(\frac{R_{\rm B}}{R_{\rm H}}\right)^{1/2}+0.13,
\label{l}
\end{equation}
where $R_{\rm B}\equiv GM_{\rm p}/c_{\rm s,PPD}^{2}$ is the Bondi radius. Here, we use $c_{\rm s,PPD}=\sqrt{k_{\rm B}T_{\rm PPD}/m_{\rm g}}$ as the isothermal sound speed around the planet, where $k_{\rm B}$, $T_{\rm PPD}$, and $m_{\rm g}$ are the Boltzmann constant, the temperature of the protoplanetary disk around the planet, and the mean molecular mass of the gas, respectively.

The heat sources determining the CPD temperature treated in the model are, the viscous heating, the irradiation from the central planet, the shocks at the surfaces of the disk by the gas accretion, and the irradiation from the PPD (see \citetalias{shi24} for the detailed formulations of the seat sources). In the case of a CPD, the viscous heating and the irradiation from PPD are the dominant heat sources. The contribution of the irradiation from the PPD is expressed as the temperature of the PPD, and we estimate the value as follows in this work. We use the midplane temperature profile of the PPD with a simplified expression for a flared disk passively heated by a central star in radiative equilibrium \citep[e.g.,][]{dull2001}:
\begin{equation}
T_{\rm PPD}=\left(\frac{\phi L_{*}}{8\pi\sigma_{\rm SB}a_{\rm p}^{2}}\right)^{1/4},
\label{TPPD}
\end{equation}
where $L_{*}$ is the stellar luminosity (59~$L_{\odot}$; \citet{Currie2022}), $\phi$ is the flaring angle (0.02 adapted by \citet{huan18}), and $\sigma_{\rm SB}$ is the Stefan--Boltzmann constant. At AB~Aur~b's location of $a_{\rm p}=93.9~{\rm au}$, $T_{\rm PPD}$ is $35~{\rm K}$. In this work, we fix the dust-to-gas surface density ratio of the CPD used for the calculation of $T$ as $Z_{\Sigma{\rm,est}}=10^{-4}$ ($r\leq r_{\rm inf}$) and $10^{-6}$ ($r>r_{\rm inf}$), which is roughly consistent with the corresponding ratio calculated in the subsequent steps of the dust evolution model, $Z_{\Sigma}=\Sigma_{\rm d}/\Sigma_{\rm g}$, where $\Sigma_{\rm d}$ is the dust surface density (see Appendix \ref{sec:opacity} for the detailed discussion).

When the gas disk is massive enough, the gravitational instability can be triggered in the disk. The condition for the occurrence of the instability is that the famous ``Toomre's Q parameter'' (but it is not a parameter in this work) is smaller than unity \citep{toomre1964},
\begin{equation}
Q_{\rm Toomre}\equiv\frac{c_{\rm s}\Omega}{\pi G\Sigma_{\rm g}},
\label{QToomre}
\end{equation}
where $c_{\rm s}=\sqrt{k_{\rm B}T/m_{\rm g}}$ and $\Omega=\sqrt{GM_{\rm p}/r^{3}}$ are the sound speed and the Keplerian frequency in the CPD. Although the original disk model produced by \citetalias{shi24} cannot treat such gravitationally unstable situations, we need to consider such situations, because the gas accretion rate of AB~Aur~b is higher than that of PDS~70~b and c and the CPD can be much more massive. Hydrodynamical simulations of PPDs show that the gas disk evolves as its $Q_{\rm Toomre}$ keeps about unity when it is massive \citep{takahashi2016}. Therefore, we correct the gas surface density of the CPD by $\Sigma_{\rm g}=c_{\rm s}\Omega/(\pi G)$ when it becomes larger than unity and recalculate the disk temperature. We note that the strength of turbulence, $\alpha$, can be distinguished into two different kinds of parameters based on the physical properties in a precise sense: the efficiency of the angular momentum transport, $\alpha_{\rm acc}$, and the diffusion strength of dust particles, $\alpha_{\rm diff}$. When the gas disk is gravitationally unstable, $\alpha_{\rm acc}$ is close to unity, but $\alpha_{\rm diff}$ should keep lower value. Since the change of $\alpha_{\rm acc}$ is already expressed by the correction of the gas surface density in our gas disk model, we do not change the value of $\alpha$ in the case of the GI disks to express the kept value of $\alpha_{\rm diff}$ in our dust evolution model. We also note that, in our 1D gas and dust model, we cannot consider the effects that spiral arms formed in gravitationally unstable disk potentially gather the dust particles, which may enhance the dust growth \citep{ric06}.

\subsection{Evolution and emission of dust} \label{sec:dust}
We then calculate how dust particles evolve in the gas disk. The dust particles are supplied to the CPD coupled with the gas inflow, and they grow to pebble size with drifting toward the planet. In the model, the radial distribution of the peak size (mass) of the dust particles ($R_{\rm d}$) and the dust surface density ($\Sigma_{\rm d}$) in the CPD is calculated by considering the evolution of dust, i.e., the radial drift, collisional growth, and fragmentation. We assume that the dust accretion rate onto the CPD, $\dot{M}_{\rm d}$, is the same with the inward mass flux of the drifting dust at the inner edge of the CPD. Then, from the conservation of mass, the radial distribution of the dust surface density is calculated, which depends on the dust inward drift speed determined by the Stokes number of the dust.

We calculate the locations of H$_{2}$O and CO$_{2}$ snowlines, $r_{\rm snow,H_{2}O}$ and $r_{\rm snow,CO_{2}}$ from the condition $P_{{\rm ev},j}=P_{j}$, where $P_{{\rm ev},j}$ and $P_{j}$ are the equilibrium vapor pressure and the partial pressure of volatile species $j$, respectively. By the Arrhenius form,
\begin{equation}
P_{{\rm ev},j}=\exp{\left(-\frac{L_{j}}{T}+A_{j}\right)}~[{\rm dyn~cm^{-2}}],
\label{Pevj}
\end{equation}
where the heat of the sublimation $L_{j}$ is $L_{\rm H_{2}O}=6070~{\rm K}$ and $L_{\rm CO_{2}}=3148~{\rm K}$, and a dimensionless constant $A_{j}$ is $A_{\rm H_{2}O}=30.86$ and $A_{\rm CO_{2}}=30.01$ \citep{yam83,bau97}. When the gas disc is well mixed in the vertical direction, the partial pressure is
\begin{equation}
P_{j}=\frac{f_{j}\Sigma_{\rm d,H_{2}O}}{\sqrt{2\pi}H_{\rm g}}\frac{k_{\rm B}T_{\rm est}}{\mu_{\rm H_{2}O}},
\label{Pj}
\end{equation}
where the surface density of water is assumed as $\Sigma_{\rm d,H_{2}O}=0.5\Sigma_{\rm d}$, and $\mu_{\rm H_{2}O}$ is the molecular mass of water. The relative abundance ($f_{j}$) of CO$_{2}$ is $f_{\rm CO_{2}}=0.1$, which refers to the observed value of comets \citep{okuzumi+16}.

The dust particles fragment rather than stick if their collision velocity is higher than a threshold called critical fragmentation velocity, $v_{\rm cr}$. In the original model of \citetalias{shi24}, the velocity is an input parameter, and its range is limited. However, recent research shows that the velocity is determined by the monomer size (radius) of the dust particles, $a_{\rm mon}$. The monomer size is highly uncertain, and it can affect the evolution of the particles \citep{okuzumi+16}. We invoke the monomer size as an input parameter and estimate the critical fragmentation velocity by
\begin{equation}
v_{\rm cr,ice}=50\left(\frac{a_{\rm mon}}{0.1~\mu{\rm m}}\right)^{-5/6}=5.2\left(\frac{a_{\rm mon}}{1.5~\mu{\rm m}}\right)^{-5/6}~[{\rm m~s^{-1}}],
\label{vcr_ice}
\end{equation}
for particles covered by water-ice and
\begin{equation}
v_{\rm cr,rock}=5\left(\frac{a_{\rm mon}}{0.1~\mu{\rm m}}\right)^{-5/6}=0.52\left(\frac{a_{\rm mon}}{1.5~\mu{\rm m}}\right)^{-5/6}~[{\rm m~s^{-1}}],
\label{vcr_rock}
\end{equation}
for rocky particles \citep{wad09,wad13,gun15}.

In addition, the fragmentation depends on the condition of the particles' surfaces. Recent works show that the particles can be covered by CO$_{2}$ mantles outside CO$_{2}$ snowline, and such CO$_{2}$-mantle particles are as fragile as rocky particles \citep{mus16a,mus16b}. Therefore, we assume the critical fragmentation velocity outside CO$_{2}$ snowline as $v_{\rm cr,CO_{2}}=v_{\rm cr,rock}$ when we consider this effect. We note that CO$_{2}$ emission has not been detected in AB~Aur system \citep{riv20,riv22}, but it has been detected in a lot of protoplanetary disks by observations including those using James Webb Space Telescope (JWST) Mid-InfraRed Instrument (MIRI) \citep{gra23}, and it is the second most common molecule after water detected in a survey of PPDs using Spitzer InfraRed-Spectrograph (IRS) \citep{pon10}. In this work, we neglect the effects of other species on the stickiness of the particles than H$_{2}$O and CO$_{2}$, since it is suspicious that there is enough amount of the other species to build sufficient mantle thickness. In summary, we consider the two cases; 1) the monomers are covered by H$_{2}$O mantles outside the H$_{2}$O snowline, and 2) the monomers are covered by H$_{2}$O mantles between the H$_{2}$O and CO$_{2}$ snowlines and covered by CO$_{2}$ mantles outside the CO$_{2}$ snowline. 

If the disk temperature is larger than $1350~{\rm K}$, we treat the rocky dust as sublimated, and we only calculate the dust distribution where the temperature is lower. We note that even in the case of AB~Aur~b with $M_{\rm p}=20~M_{\rm J}$ (see Section \ref{sec:extincstion}), the temperature is higher than $1350~{\rm K}$ only at the inner part of the disk, showing that the sublimation of rocky dust does not affect our results.

Finally, we calculate the total flux density of the dust thermal emission from the entire CPD at wavelength, $\lambda$. We integrate the flux from each orbit,
\begin{equation}
F_{{\rm d},\lambda}=\frac{2\pi\cos{i}}{d^{2}}\int^{r_{\rm inf}}_{r_{\rm in}}\left\{1-\exp\left(-\frac{\tau_{\lambda}}{\cos{i}}\right)\right\}B_{\nu}rdr,
\label{Fdlambda}
\end{equation}
where $i$, $r_{\rm inf}$, $r_{\rm in}$, $\tau_{\lambda}$, and $B_{\nu}$ are the inclination of the CPD (assumed as the same with that of PPD), the location of the outer edge of the dust (and gas) inflow region, the inner edge of the CPD (calculated under the assumption of magnetospheric accretion), the optical depth at $\lambda$, and the Planck function. The optical depth is
\begin{equation}
\tau_{\lambda}=\int^{R_{\rm d}}_{a_{\rm mon}}\Sigma_{{\rm d},a}\kappa_{\rm abs}~da,
\label{tau}
\end{equation}
where $\kappa_{\rm abs}$ is the absorption mass opacity for $\lambda$ modeled by \citet{kat14} and \citet{bir18}. The opacity depends on the dust size (see Section 2.3 of \citetalias{shi24} for the detailed explanations). Here, the surface density of dust with size (radius) of $a$, $\Sigma_{{\rm d},a}$, is calculated from $\Sigma_{\rm d}$ assuming a typical size frequency distribution; $dN/da\propto a^{-3.5}$ \citep[e.g.,][]{doi23}. The maximum radius of dust is corresponding to the peak mass size $R_{\rm d}$ (\citetalias{shi24}), and we assume that the minimum radius is the monomer radius, $a_{\rm mon}$. We do not change the opacity model depending on the surface conditions of dust (i.e., covered by water/CO$_{2}$ ice or not) and note that the assumption potentially affects the results.

We note that this model does not consider any CPD substructures formed by potential large moons, which could make the total flux density of the dust emission from the CPD stronger by forming such as dust rings as seen in protoplanetary disks.

\subsection{Properties and parameters} \label{sec:parameters}
In this work, we investigate the evolution and emission of the dust in the CPD of AB~Aur~b as a case study. Table \ref{tab:properties} shows the properties of AB~Aur~b used in this work. \citetalias{Currie2022} reported detection of near-infrared continuum and H$\alpha$ line emission at the location of $93.9~{\rm au}$ from AB~Aur by using Subaru/SCExAO and the Hubble Space Telescope (HST) high-contrast imaging observations. AB~Aur is a $4.1~{\rm Myr}$ old Herbig star at $d=155.9~{\rm pc}$ with the stellar mass of $M_{*}=2.4~M_\sun$ and $T_{\rm eff}=9000~{\rm K}$ \citep{Guzman2021a}, which is one of the most intensively studied intermediate mass star. Its system is also unique; multiple spiral arms have been reported in a variety of infrared and (sub)millimeter wavelength observations \citep[e.g.,][]{fuk04,has11,tang12}. Especially, \citet{Tang2017} has identified two spiral arms in $^{12}$CO line emission by ALMA observations and argues that tidal disturbance caused by a potential companion (planet) can explain the morphology, where the predicted location of the companion is near to the near-infrared continuum and H$\alpha$ line emission spot reported in \citetalias{Currie2022} (see also Appendix \ref{sec:previous}). \citetalias{Currie2022} also obtained the planetary radius and planetary effective temperature of AB~Aur~b by searching the best-fit parameters of a model reproducing the observed planetary spectral energy distribution (SED) of near-infrared, $R_{\rm p}=2.75~R_{\rm J}$ and $T_{\rm p,eff}=2200~{\rm K}$, respectively.

We also consider the contribution of the irradiation from the surrounding PPD to the temperature of the CPD by considering the midplane temperature of the PPD at the orbital location of the planet $T_{\rm PPD}=35~{\rm K}$ (see Section \ref{sec:gasdisk}).

Table \ref{tab:parameters} shows the parameters of AB~Aur~b in each section of this work. In Section \ref{sec:results}, we use the values of the planet mass and the gas accretion rate of AB~Aur~b obtained by the best-fit of the SED model to the observed near-infrared; $M_{\rm p}=9~M_{\rm J}$ and $\dot{M}_{\rm g}=1.1\times10^{-6}~M_{\rm J}~{\rm yr}^{-1}$ (\citetalias{Currie2022}). In Sections \ref{sec:extincstion} and \ref{sec:band7}, however, we investigate the cases where the values obtained in \citetalias{Currie2022} are underestimate due to the effects of the extinction by small grains. We obtain corrected values; $M_{\rm p}=20~M_{\rm J}$, and $\dot{M}_{\rm g}=2.2\times10^{-6}$ or $5.3\times10^{-6}~M_{\rm J}~{\rm yr}^{-1}$ (see Section \ref{sec:extincstion} for the details). We refer to such case as \texttt{Extinction}.

The strength of turbulence in the CPD, $\alpha$, is another important parameter, but it is highly unknown. Theoretical prediction argues that magnetorotational instability does not likely occur in CPDs \citep{fuj14,tur14}, but tidal effects from the star may make spiral shocks in CPDs, and they could transport angular momentum and drive gas accretion of the CPD \citep{zhu16}. In this work, we widely change the value of $\alpha$ from $10^{-6}$ to $10^{-2}$.

Considering the solar composition and the possibility of the dust filtering at the edge of the gas gap created by the planet, the dust-to-gas mass ratio in the gas inflow onto the CPD, $x$, should be lower than $0.01$ \citep[e.g.,][]{hom20}. On the other hand, numerical simulations show that small particles can penetrate into gaps and brought to the vicinity of the planets with gas \citep[e.g.,][]{szu22}. Recently, \citet{mae24} investigated delivery of dust by local simulations. They found that the dust-to-gas mass ratio of the inflow to the CPD is about 0.001 when the dust-to-gas surface density ratio is 0.01 and the dust is blown up to high altitude enough by such as turbulence at the root of the gas inflow. They only investigated the situations where the ratio of the Hill radius of a gas accreting planet to the scale height of the PPD is from 0.5 to 1.36. In the case of AB~Aur~b with $M_{\rm p}=9~M_{\rm J}$, the Hill radius to scale height ratio is about 1.4, which is almost the same with the largest value of the parameter investigated in the work. On the other hand, in the case of AB~Aur~b with $M_{\rm p}=20~M_{\rm J}$ (see Section \ref{sec:extincstion}) the ratios are outside the investigated parameter range. Therefore, in this work, we do not fix the dust-to-gas mass ratio in the inflow as $x=0.001$ but change the value from $0.0001$ to $0.01$ and treat $x=0.001$ as just a typical value.

We assume the monomer radius of the dust particles is $a_{\rm mon}=0.1~\mu{\rm m}$ but also consider the cases where the monomers are large, $a_{\rm mon}=1.5~\mu{\rm m}$, which makes the dust particles fragile. In any calculations of this work, we consider that the monomers are covered by H$_{2}$O mantles outside the H$_{2}$O snowline. We then consider the cases where the monomers are additionally covered by CO$_{2}$ mantles outside the CO$_{2}$ snowline, which also makes the particles fragile (see Section \ref{sec:dust} for the details). We then investigate the cases where the both effects are considered. We refer to theses cases as \texttt{Large monomer}, \texttt{CO$_{2}$ mantle}, and \texttt{Large monomer + CO$_{2}$ mantle}, respectively, and the case we do not consider these effects as \texttt{Fiducial}.

We investigate the flux density of the dust emission from the CPD at ALMA Band 6, the wavelength of $\lambda=1.3~{\rm mm}$ to compare our predictions with the previous observation of AB~Aur by \citet{Tang2017}. In Section \ref{sec:band7}, we show the cases at ALMA Band 7, $\lambda=855~\mu{\rm m}$, to compare the predictions with the previous observation of PDS~70~c and discuss future observations.

\begin{table}[tbp]
\caption{Properties of AB~Aur and AB~Aur~b}
\label{tab:properties}
\begin{tabular}{lll}
\hline
\hline
Properties & Symbol & Value \\
\hline
Stellar mass & $M_{*}$ & $2.4~M_{\odot}$ \\
Stellar luminosity & $L_{*}$ & $59~L_{\odot}$ \\
Distance to the system & $d$ & $155.9~{\rm pc}$ \\
Inclination of PPD and CPD & $i$ & $42.6\degr$ \\
\hline
Planetary orbit & $a_{\rm p}$ & $93.9~{\rm au}$ \\
Planetary radius & $R_{\rm p}$ & $2.75~R_{\rm J}$ \\
Planetary effective temperature & $T_{\rm p,eff}$ & $2200~{\rm K}$ \\
Temperature of PPD & $T_{\rm PPD}$ & $35~{\rm K}$ \\
\hline
\end{tabular}
\end{table}

\begin{table}[tbp]
\centering
\caption{Parameters of AB~Aur~b in each section}
\label{tab:parameters}
\begin{tabular}{lll}
\hline
Parameters & Symbol & Value \\
\hline
Section \ref{sec:distribution} & & \\
\hline
Planet mass & $M_{\rm p}$ & $9~M_{\rm J}$\\
Gas accretion rate & $\dot{M}_{\rm g}$ & $1.1\times10^{-6}~M_{\rm J}~{\rm yr}^{-1}$ \\
Strength of turbulence in CPD & $\alpha$ & $10^{-3},10^{-4},10^{-5}$ \\
Dust-to-gas mass ratio in inflow & $x$ & $0.001,0.01$ \\
Monomer radius & $a_{\rm mon}$ & $0.1~\mu{\rm m}$ \\
Wavelength & $\lambda$ & $1.3~{\rm mm}$ \\
\hline
Section \ref{sec:monomer} & & \\
\hline
& $M_{\rm p}$ & $9~M_{\rm J}$ \\
& $\dot{M}_{\rm g}$ & $1.1\times10^{-6}~M_{\rm J}~{\rm yr}^{-1}$ \\
& $\alpha$ & $3\times10^{-3}$ \\
& $x$ & $0.001$ \\
& $a_{\rm mon}$ & $0.1,1.5~\mu{\rm m}$ \\
& $\lambda$ & $1.3~{\rm mm}$ \\
\hline
Section \ref{sec:extincstion} & & \\
\hline
& $M_{\rm p}$ & $20~M_{\rm J}$ \\
& $\dot{M}_{\rm g}$ & $2.2\times10^{-6}, 8.9\times10^{-6}$ \\
& & $M_{\rm J}~{\rm yr}^{-1}$ \\
& $\alpha$ & $10^{-6}-10^{-2}$ \\
& $x$ & $0.0001-0.01$ \\
& $a_{\rm mon}$ & $0.1,1.5~\mu{\rm m}$ \\
& $\lambda$ & $1.3~{\rm mm}$ \\
\hline
Section \ref{sec:band7} & & \\
\hline
& $M_{\rm p}$ & $9,20~M_{\rm J}$ \\
& $\dot{M}_{\rm g}$ & $1.1\times10^{-6}, 2.2\times10^{-6},$ \\
& & $8.9\times10^{-6}~M_{\rm J}~{\rm yr}^{-1}$ \\
& $\alpha$ & $10^{-6}-10^{-2}$ \\
& $x$ & $0.0001-0.01$ \\
& $a_{\rm mon}$ & $0.1,1.5~\mu{\rm m}$ \\
& $\lambda$ & $855~\mu{\rm m}$ \\
\hline
\end{tabular}
\end{table}


\section{Results} \label{sec:results}
In this section, we show the predictions of the flux density of the dust emission from the potential CPD of AB~Aur~b, $F_{{\rm d}, \lambda}$, as a case study. However, the tends and features of the results obtained in this section, such as the parameter dependence and the effects of the conditions of the monomers, may also be applied to the potential CPDs of other gas accreting planets.

\subsection{Gas and dust distribution in the CPD} \label{sec:distribution}
\begin{figure}[tbp]
\center
\includegraphics[width=\linewidth]{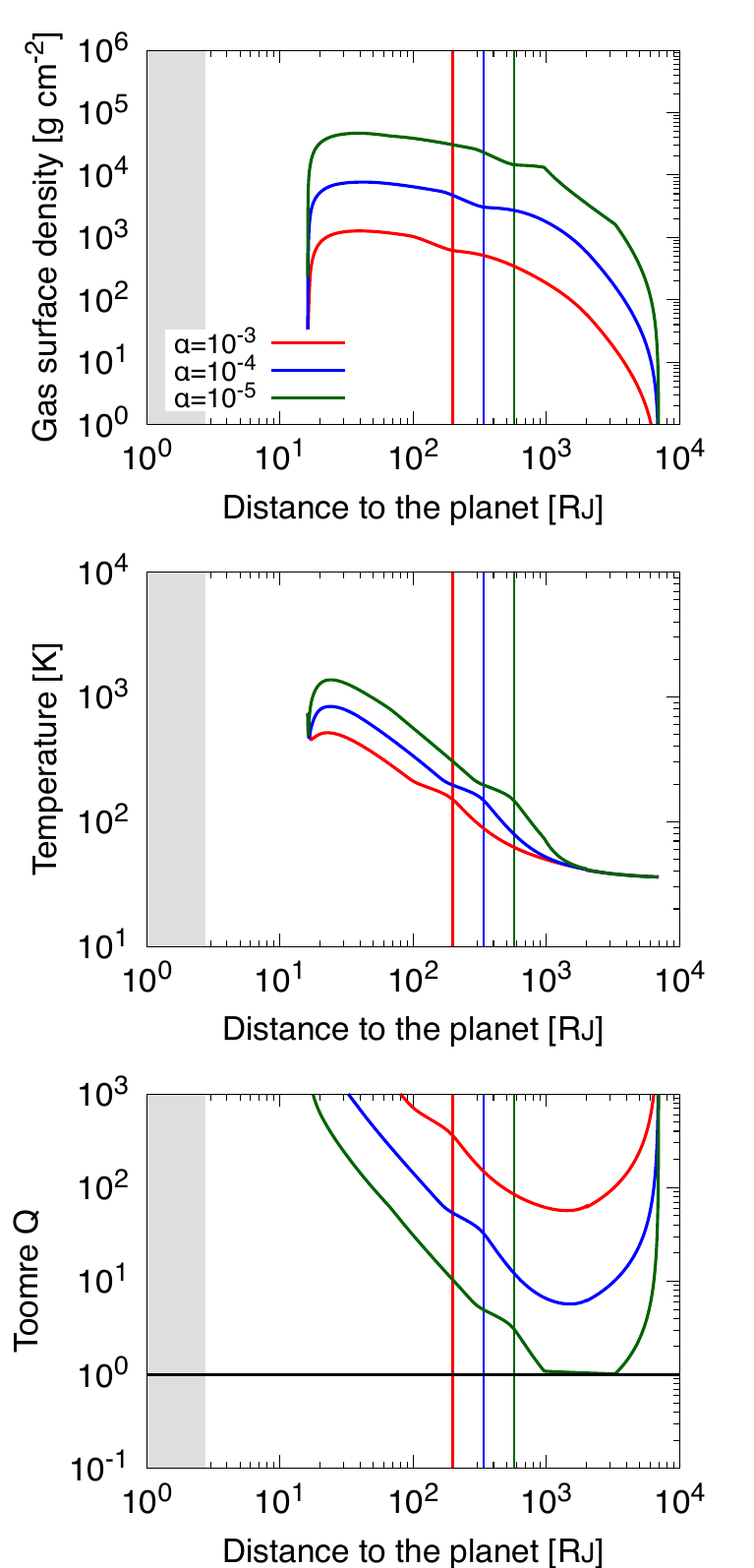}
\caption{Radial distribution of the gas surface density, temperature, and Toomre Q parameter of the CPD of AB~Aur~B. The red, blue, and green curves represent the profiles with $\alpha=10^{-3}$, $10^{-4}$, and $10^{-5}$, respectively. The vertical lines are the H$_{2}$O snow lines. The black horizontal line in the bottom panel is $Q_{\rm Toomre}=1$. Shaded gray regions represent the planetary surface (atmosphere), $r\leq2.75~R_{\rm J}$ (\citetalias{Currie2022}).} \label{fig:gas-distribution}
\end{figure}

We first show the predicted gas and dust radial distribution in the potential CPD of AB~Aur~b. The gas profiles mainly depend on the planet mass ($M_{\rm p}$), the gas accretion rate ($\dot{M}_{\rm g}$), and the strength of turbulence in the CPD ($\alpha$). The top panel of Fig. \ref{fig:gas-distribution} represents the gas surface density of the disk ($\Sigma_{\rm g}$). When $M_{\rm p}$ and $\dot{M}_{\rm g}$ are fixed, the gas surface density of a viscous accretion disks is about proportional to $1/\alpha$. As a result, the gas surface density is large as $\alpha$ is small. The slope of the outer part of the CPD when $\alpha=10^{-5}$ is different from those of the other two cases, because the gas disk is gravitationally unstable at the part (bottom panel). The middle panel represents the midplane disk temperature. The temperature of the outer part of the CPD is determined by the temperature of PPD, $T_{\rm PPD}=35~{\rm K}$. The heat source of the other part is dominated by the viscous heating, resulting in the temperature dependent on $\alpha$ and $x$. The temperature is large as $\alpha$ is small, because the dust-to-gas surface density ratio of the disk calculated for the opacity used in the calculation of the temperature, $Z_{\Sigma,{\rm est}}$, is fixed. When $\alpha$ is small, $\Sigma_{\rm g}$ and $\Sigma_{\rm d}$ are large, and the optical depth is large, resulting in high temperature. As a result, the locations of H$_{2}$O snowlines shift outwards as $\alpha$ is small. The bottom panel of Fig. \ref{fig:gas-distribution} represents the profiles of Toomre Q parameter of the disk. When $\alpha=10^{-5}$, the Toomre Q parameter reaches unity at the outer part of the disk due to the high gas surface density, making that part of the CPD gravitationally unstable.

\begin{figure*}[tbp]
\center
\includegraphics[width=\linewidth]{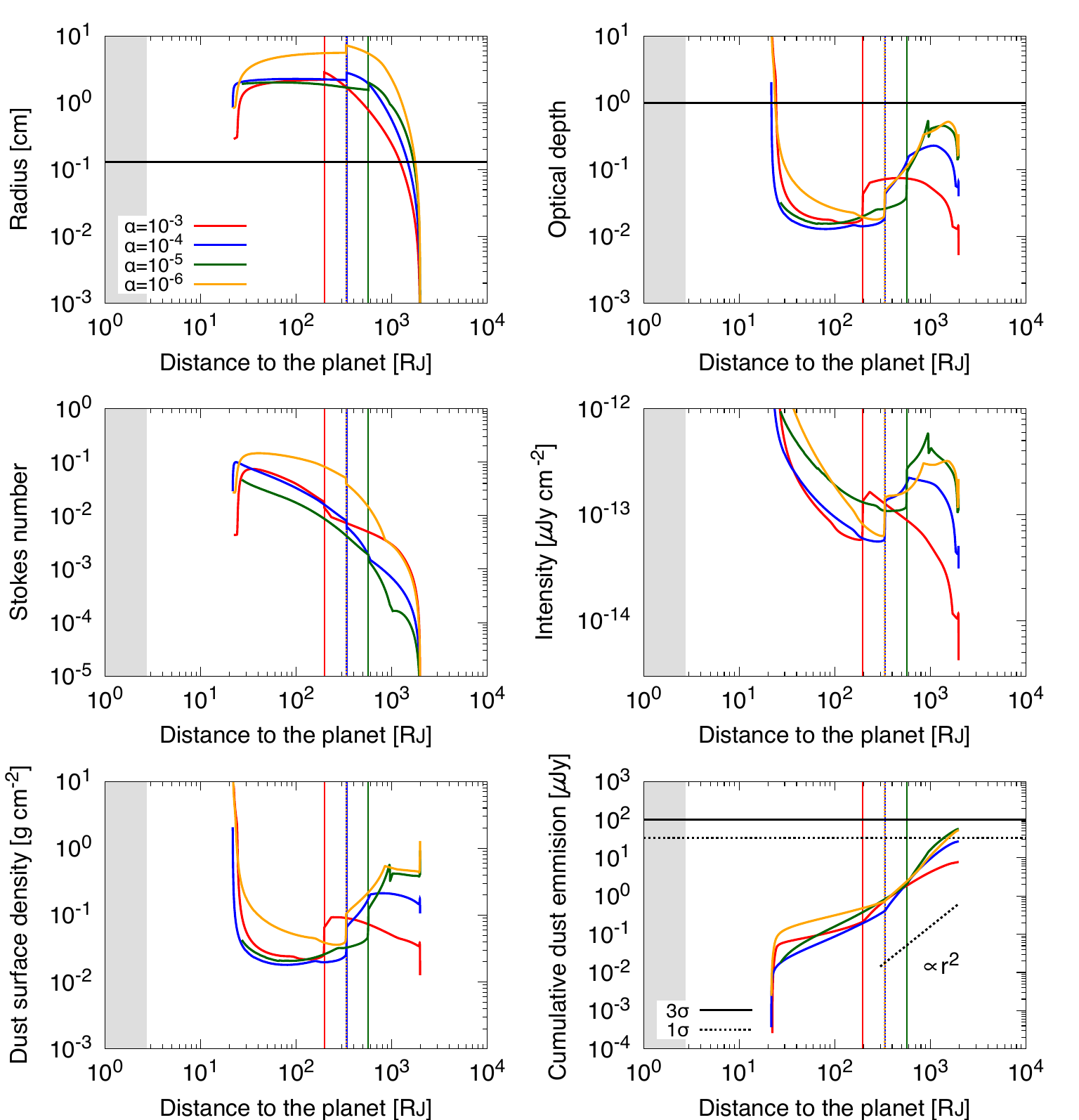}
\caption{Radial distribution of the dust properties in the CPD of AB~Aur~B. The top left, middle left, bottom left, top right, middle right, and bottom right panels represent the dust particle radius, Stokes number, dust surface density, optical depth, intensity, and cumulative dust emission (flux density), respectively. The assumed wavelength in the right three panels is $\lambda=1.3~{\rm mm}$. The orange curves and vertical dotted lines represent the profiles with $\alpha=10^{-4}$ and $x=0.01$. The other color of the curves, the vertical lines, and the shaded gray regions have the same meanings with those in Fig. \ref{fig:gas-distribution}. The horizontal lines in the top left and top right panels represent the observational wavelength and $\tau_{\rm \lambda}=1$, respectively. The solid and dotted horizontal lines in the bottom right panel represent the noise levels of $3~\sigma=99~\mu{\rm Jy}$ and $1~\sigma~\mu{\rm Jy}=33~\mu{\rm Jy}$ of the previous observation by \citet{Tang2017} (see Appendix \ref{sec:previous}), respectively.}
\label{fig:dust-distribution}
\end{figure*}

Figure \ref{fig:dust-distribution} represents the steady-state radial distribution of the dust particles. The top left panel shows that the dust grows to cm-size and drifts inward. The steps around $r=200-500~R_{\rm J}$ are formed by the H$_{2}$O snowlines. When $x$ is large (orange), the sizes of the particles and the dust surface density are large. On the other hand, the difference of $\alpha$ does not affect the size so much, because small $\alpha$ makes $\Sigma_{\rm g}$ large and the drift speed of particles fast, but the dust also settles onto the midplane when $\alpha$ is small, resulting in efficient dust growth. As a result, these effects are almost canceled out, the $\alpha$ dependence is weak on the dust size. We note that, here, the $\alpha$ parameter when the CPD is gravitationally unstable (i.e., $\alpha\lesssim10^{-5}$) does not express the transport of the angular momentum (which should be close to unity due to the GI) but the strength of the dust diffusion. The sizes of particles inside the snowlines are smaller than those of the outside, because the internal density of rocky dust is higher than that of icy dust. The top left panel also shows that the size of the particles is getting small as they move inward inside the H$_{2}$O snowlines, (red and orange), because rocky particles are more fragile than icy particles (Eqs. (\ref{vcr_ice}) and (\ref{vcr_rock})). This fragmentation occurs when the turbulence is strong and the collision speed of the particles is fast (orange), or the dust-to-gas mass ratio in the inflow is large and the dust density on the midplane is large (red).

One important characteristic of the dust in the CPD is that the dust particles drift inward significantly and the dust surface density is much lower than the gas surface density; the dust-to-gas surface density ratio at $r=100~R_{\rm J}$ is $\sim10^{-6}-10^{-5}$ (see also Appendix \ref{sec:opacity}). This quite low dust surface density makes the disk optically thin and the flux density of the dust thermal emission lower.

The changes of the slopes around $r=800-1000~R_{\rm J}$ in the Stokes number (left middle panel, green and orange) are the locations where the particles enter the Stokes regime from the Epstein regime. Outside the snowline, the dust surface density decreases as the particles drift inward (left bottom panel), because their Stokes number grows large, and the drift speed becomes faster. Since the dust mass flux inside the snowlines is half of the outside, the dust surface density inside the snowlines is about half of the outside. The top right panel of Fig. \ref{fig:dust-distribution} shows that the whole CPD of AB~Aur~b is optically thin at $\lambda=1.3~{\rm mm}$, and the opacity is higher at the outer part. As a result, the intensity of the dust thermal emission is higher at the outer part (right middle). The right bottom panel then shows that the $r$ dependence of the cumulative dust emission (cumulative from $r_{\rm in}$ to $r$) is stronger than $r^{2}$ except for the case with $\alpha=10^{-3}$, where it is proportional to the surface area of the dust {-containing region of the CPD} (i.e., $r^{2}$) if the intensity is radially uniform. This results show that the angular momentum of the gas inflow is an important factor determining the total density flux of the dust emission from the CPD.

When $\alpha=10^{-3}$ (red curve), due to the smaller gas surface density, the Stokes number of dust is high at the outer part of the disk compared to the cases with $\alpha=10^{-4}$ and $10^{-5}$ (blue and green), and so the drift speed is high, resulting in lower dust surface density at the outer part of the disk (bottom left). Therefore, the intensity of the dust emission at the outer part is lower (middle right), and the slope of the cumulative dust emission is shallower (bottom right) than those in Fig. \ref{fig:dust-distribution}. As a result, the flux density of the dust emission from the entire CPD is weaker, and its $r_{\rm inf}$ dependence should also be weaker than those with $\alpha=10^{-4}$ and $10^{-5}$.

The $x$ dependence on the location of the H$_{2}$O snowline is very weak. When $x=0.001$ (blue), the location is $r=338~R_{\rm J}$, and when $x=0.01$, it is $r=334~R_{\rm J}$ (orange). The locations of the snowlines mainly depend on the temperature of the disk (see Eqs. (\ref{Pevj}) and (\ref{Pj})).

\subsection{Effects of the conditions of the monomers} \label{sec:monomer}
We then investigate the effects of the conditions of the monomers. We consider the cases where the dust particles are more fragile; the monomers are large ($a_{\rm mon}=1.5~\mu{\rm m}$; \texttt{Large monomer}), covered by CO$_{2}$ mantles outside the CO$_{2}$ snowlines (\texttt{CO$_{2}$ mantle}), or both (\texttt{Large monomer + CO$_{2}$ mantle}) (see Section \ref{sec:dust}). In this section, we assume stronger turbulence in the CPD, $\alpha=3\times10^{-3}$, than the values assumed in Section \ref{sec:distribution} to show the effects of the large monomers and CO$_{2}$ mantles clearly.

\begin{figure*}[tbp]
\center
\includegraphics[width=\linewidth]{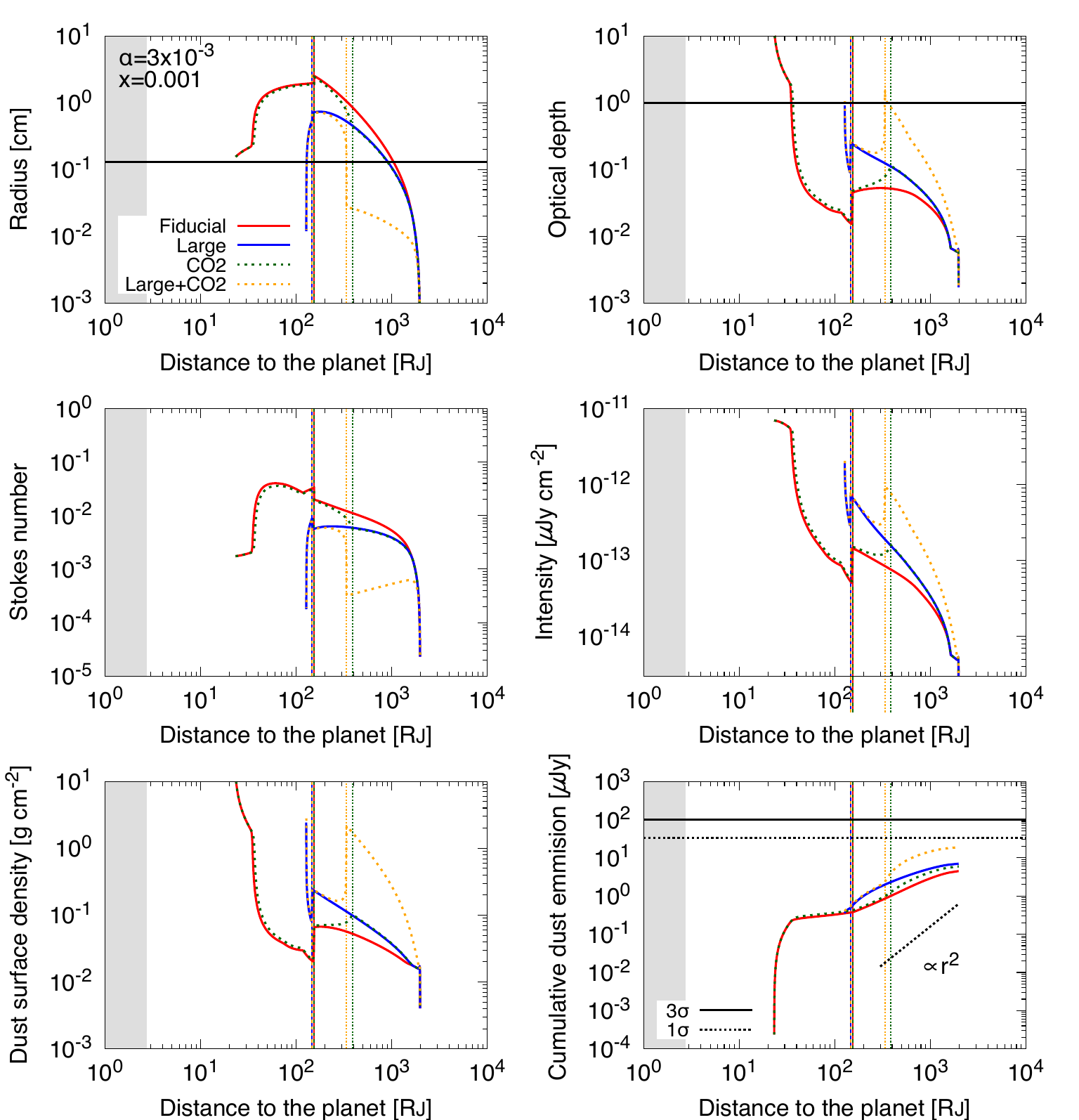}
\caption{Same as Fig. \ref{fig:dust-distribution} but $\alpha=3\times10^{-3}$, $x=0.001$, and the different monomer conditions. The red, blue, green, and orange curves represent the cases of \texttt{Fiducial}, \texttt{Large monomer}, \texttt{CO$_{2}$ mantles}, and \texttt{Large monomer + CO$_{2}$ mantles}, respectively. The vertical solid/dashed and dotted lines are the corresponding H$_{2}$O and CO$_{2}$ snowlines, respectively.}
\label{fig:conditions}
\end{figure*}

Figure \ref{fig:conditions} shows the radial distribution of dust properties in the CPD of AB~Aur~b with $\alpha=3\times10^{-3}$. When the monomers are large (\texttt{Large monomer}, blue curves), the particles are fragile, and the radius of the particles are smaller than \texttt{Fiducial} (red) outside the H$_{2}$O snowline (top left panel). When the dust size and the Stokes number is small (middle left), the drift speed is slow, and the surface density is large (bottom left). Therefore, the optical depth and the intensity of dust emission is higher than \texttt{Fiducial} (top and middle right), resulting in stronger dust emission from the CPD (bottom right). When the monomers are covered by CO$_{2}$ mantles (\texttt{CO$_{2}$ mantle}, green), the critical fragmentation velocity is slower than that of the water-ice particles. Therefore, the evolution track and the intensity outside the CO$_{2}$ snowline are almost the same as those of \texttt{Large monomer}. As a result, the total flux density of the dust emission is larger than \texttt{Fiducial} (bottom right), because it is dominated by the outer region of the disk, outside the CO$_{2}$ snowline. Inside the CO$_{2}$ snowline, the particles grow faster and its size reaches to the \texttt{Fiducial} case. When the monomer is large and covered by CO$_{2}$ (\texttt{Large monomer + CO$_{2}$ mantle}, orange), the critical velocity is about 100 times slower than \texttt{Fiducial}, which makes the size and the Stokes number of the dust much smaller than the \texttt{Fiducial} case (top and middle left, orange). As a result,the dust surface density is larger, making the intensity and the total flux density of the dust emission higher than \texttt{Fiducial} (middle and bottom right), although the size of the dust is smaller then the observation wavelength (top left).

\subsection{Expected dust emission at ALMA Band 6} \label{sec:band6}
We then change the two uncertain parameters in broad ranges, $10^{-6}\leq\alpha\leq10^{-2}$ and $0.0001\leq x\leq0.01$, and compare the predictions with the previous ALMA observation by \citet{Tang2017} (see Appendix \ref{sec:previous}). Figure \ref{fig:prediction} shows the expected flux density of the dust continuum emission from the CPD of AB~Aur~b at Band 6 ($\lambda=1.3~{\rm mm}$). The upper left panel shows that the expected dust emission in the \texttt{Fiducial} case is lower than the $3~\sigma=99~\mu{\rm Jy}$ of the previous observation (the green solid line) with any $\alpha$ when the dust-to-gas mass ratio in the inflow is the typical value $x=0.001$ (white curve). The expected emission with $x=0.001$ is also lower than $1~\sigma=33~\mu{\rm Jy}$ when $\alpha>10^{-4}$. The positive correlation between $x$ and $F_{\rm d}$ is because the dust surface density is lager as the dust accretion rate onto the CPD is larger. The negative correlation between $\alpha$ and $F_{\rm d}$ in $\alpha\gtrsim10^{-5}$ is explained as follows (see also Section \ref{sec:distribution}). When the turbulence is strong, the gas surface density is low ($\Sigma_{\rm g}$ is about proportional to $\dot{M}_{\rm g}/\alpha$ in viscous accretion disks). The dust particles have larger Stokes number in low gas (surface) density, resulting in their faster drift and lower surface density. As a result, $F_{\rm d}$ is small when $\alpha$ is large. The transition of the trend at $\alpha\lesssim10^{-5}$ is cased by that the Toomre Q is lower than unity, which makes $\Sigma_{\rm g}$ constant with respect to $\alpha$ (see Section \ref{sec:gasdisk}). Since the dust particles can grow larger when $\alpha$ is smaller, the larger radius means the larger Stokes number when $\Sigma_{\rm g}$ is constant, resulting in faster radial drift and lower surface density of dust. As a result, $F_{{\rm d},\lambda}$ is smaller as $\alpha$ is smaller when the gas disk is gravitationally unstable.

The upper right panel shows that the $\alpha$ dependence shown in the \texttt{Fiducial} case changes when the monomer size is large (\texttt{Large monomer}). The particles are more fragile, and it makes the total flux density higher as we explained in Section \ref{sec:monomer}. This effect is strong as $\alpha$ is large, because the collision velocity of the particles is determined by the strength of turbulence (when it is strong enough). This trend is also case when we assume the monomers are covered by CO$_{2}$ mantles outside the CO$_{2}$ snowline (lower left panel; \texttt{CO$_{2}$ mantle}). When the monomers are both large and covered by CO$_{2}$ mantles (\texttt{Large monomer + CO$_{2}$ mantle}), the flux density with large $\alpha$ is much higher, so that its $\alpha$ dependence becomes weak. However, the expected dust emission is still lower than the $3~\sigma=99~\mu{\rm Jy}$ of the previous observation \citep{Tang2017}, which can be the reason of the non-detection. The panel also shows that only the observations with the noise levels of $\sigma\lesssim10~\mu{\rm Jy}$ can detect the CPD at Band 6 with $3~\sigma$ when $x=0.001$ (typical value).

\begin{figure*}[tbp]
\centering
\includegraphics[width=0.49\linewidth]{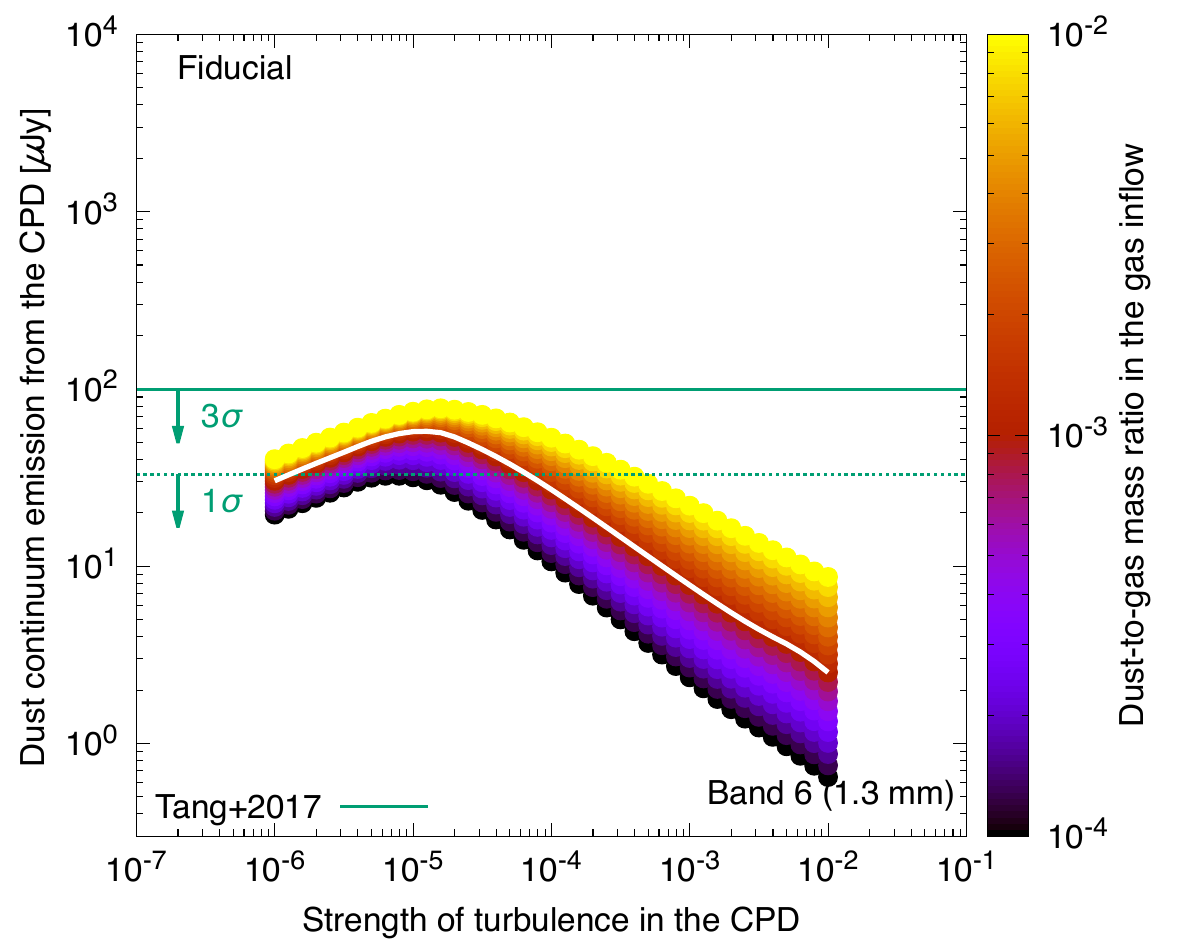}
\includegraphics[width=0.49\linewidth]{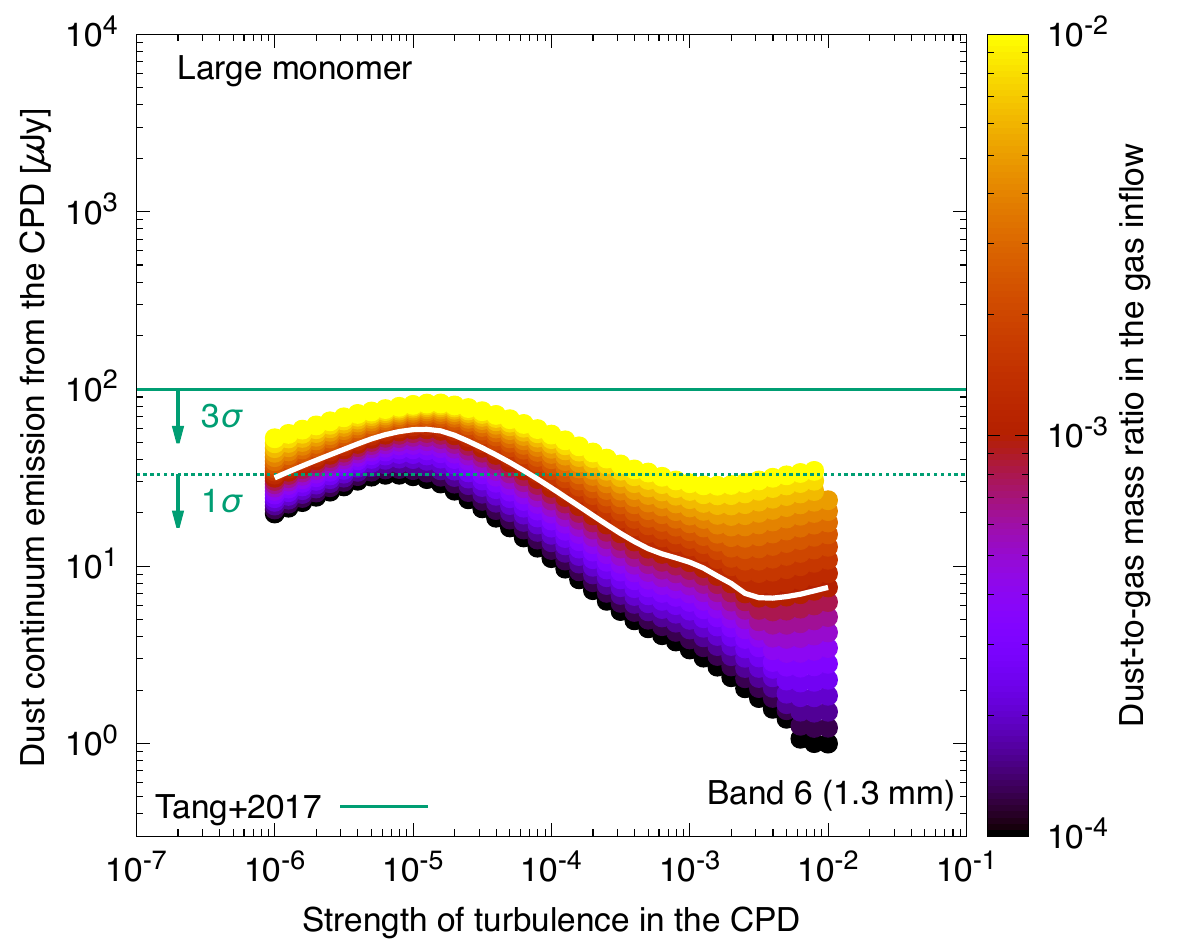}
\includegraphics[width=0.49\linewidth]{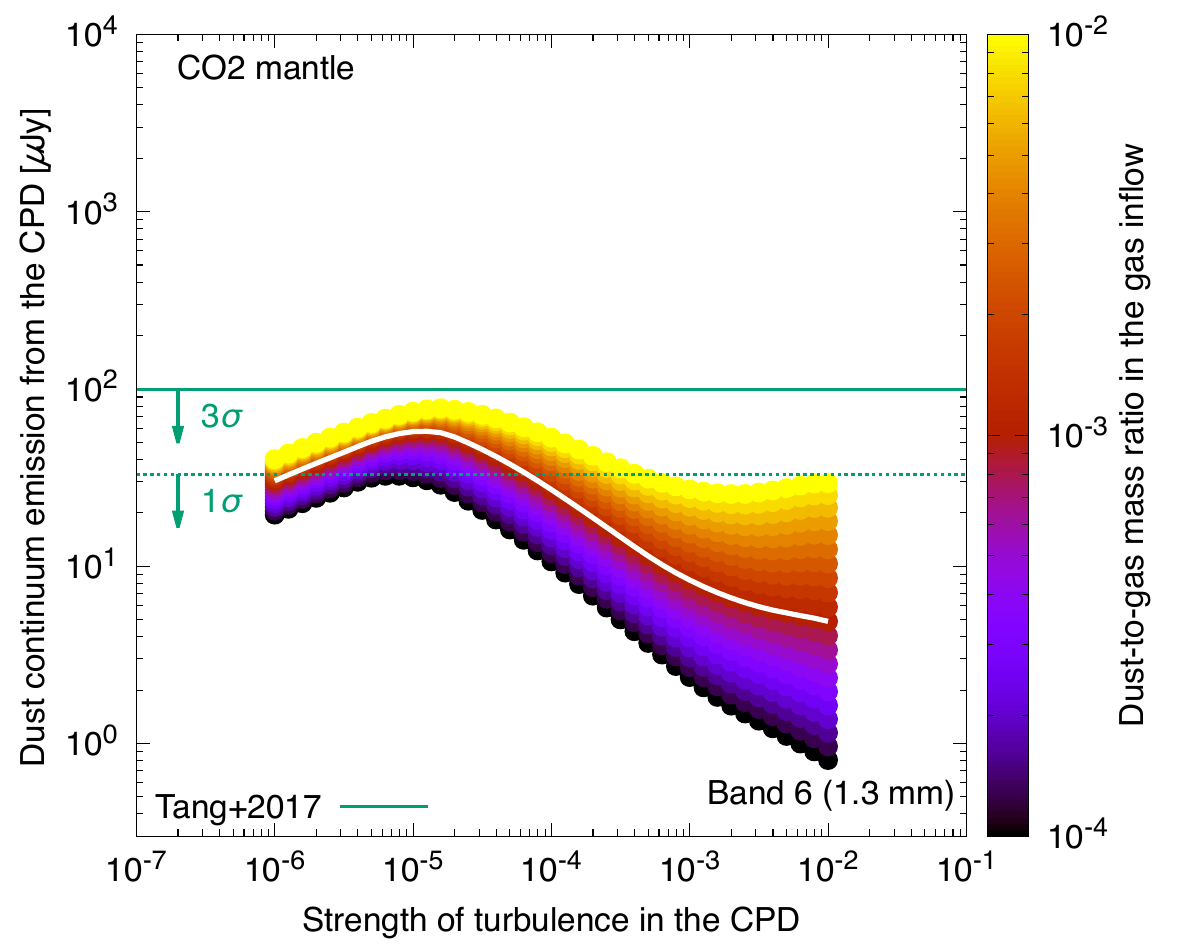}
\includegraphics[width=0.49\linewidth]{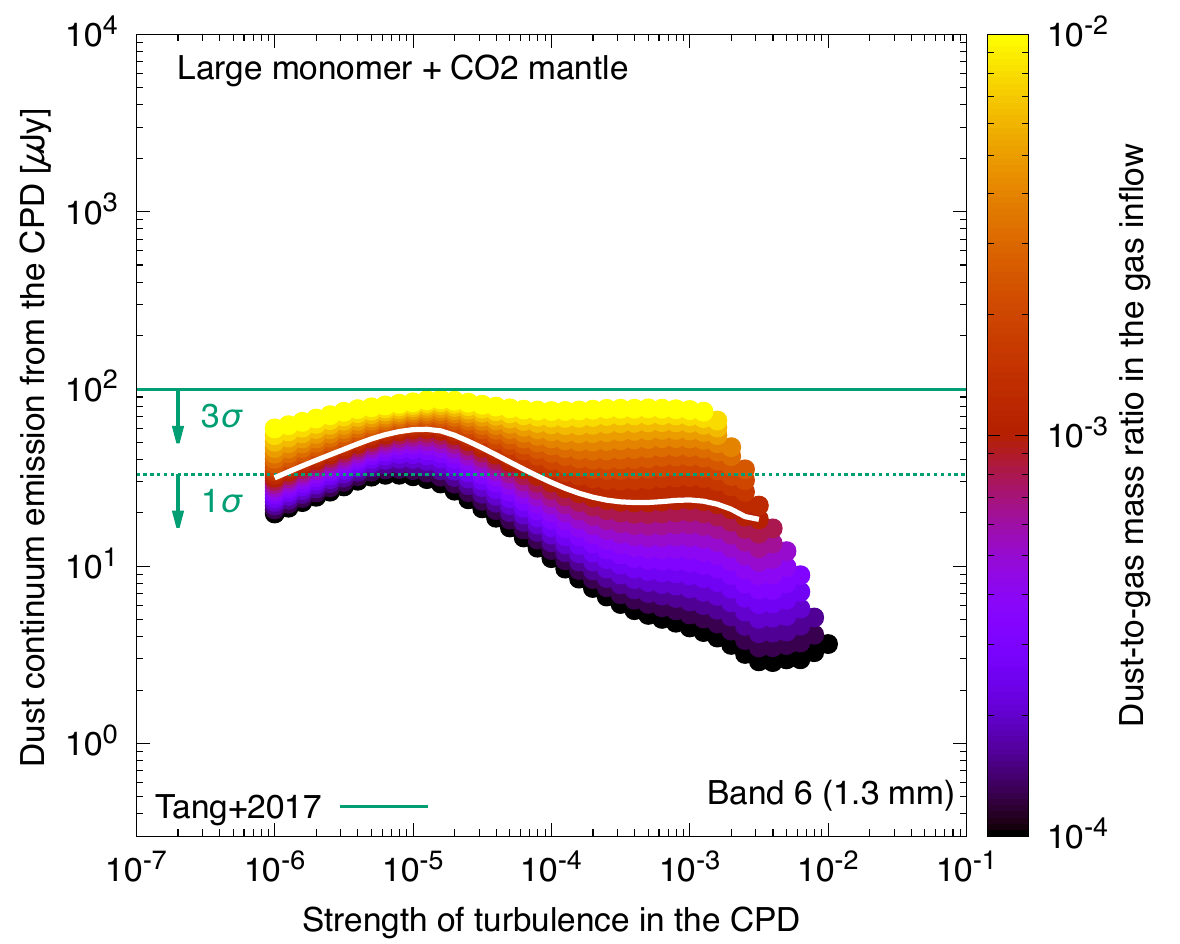}
\caption{Expected flux density of the dust continuum emission from the CPD of AB~Aur~b at ALMA Band 6 ($\lambda=1.3~{\rm mm}$). The color of the circles represent the dust-to-gas mass ratio $x$ in the gas inflow onto the CPD, and the white curves are the cases where $x=0.001$. The green solid and dotted horizontal lines represent the noise levels of $3~\sigma=99~\mu{\rm Jy}$ and $1~\sigma~\mu{\rm Jy}=33~\mu{\rm Jy}$ of the previous observation by \citet{Tang2017} (see Appendix \ref{sec:previous}), respectively. The upper left, upper right, lower left, and lower right panels represent \texttt{Fiducial}, \texttt{Large monomer}, \texttt{CO$_{2}$ mantles}, and \texttt{Large monomer + CO$_{2}$ mantles}, respectively. We note that some results with $\alpha\sqrt{x}\gtrsim10^{-4}$ are not presented in the lower right panel because of numerical reasons.
} \label{fig:prediction}
\end{figure*}

\section{Discussion}
\subsection{Effects of the reduction of the observed near-infrared emission due to the extinction by small grains} \label{sec:extincstion}
One possible reason of rare direct detection of embedded accreting planets in near-infrared emission is the extinction caused by micron-sized dust grains covering the planets \citep{Hashimoto2020,Marleau2022a,ala24}. In the previous sections, we have used the planet mass and the gas accretion rate obtained in \citetalias{Currie2022}, where the extinction is not considered. However, \citetalias{Currie2022} also shows that the optical depth at the planet location in H band is about $\tau_{\rm H}=1.3$ when their PPD model assuming the presence of an embedded planet (i.e., AB~Aur~b) can reproduce the distribution of the scattered light brightness over the whole PPD\footnote{\citetalias{Currie2022} shows the observed scattered light brightness can also be reproduced with $\tau_{\rm H}=0.25-2$, but we here use the value of $\tau_{\rm H}=1.3$ as a representative value in this work.}. In addition, although \citetalias{Currie2022} estimated the planet mass and the gas accretion rate from the best fit of a SED model by \citet{zhu2015} to the observations, the model only depends on the contribution from the planet atmosphere at H band; that contribution does not depend on the gas accretion rate but on the planet mass \citep{spie12}. Therefore, we here estimate the planet mass and the gas accretion rate from the observed absolute magnitude at H band and the H$\alpha$ luminosity, respectively.

The optical depth at H band, $\tau_{\rm H}=1.3$, means that the actual absolute magnitude of the planet is $1.3$ times higher than the observed value. According to an evolutionary model for very low mass stars and brown dwarfs, this upward revision of the absolute magnitude roughly corresponds to the upward revision of the planet mass from $9$ to $20~M_{\rm J}$ \citep{cha00}.

The observed H$\alpha$ luminosity is $L_{\rm H\alpha}=1.4\pm0.4\times10^{-5}~L_{\odot}$, where $L_{\odot}=3.83\times10^{26}~{\rm erg~s}^{-1}$ is the solar luminosity \citep{zho22}\footnote{\citet{zho22} obtains $L_{\rm H\alpha}=2.2\pm0.7~L_{\odot}$ considering the extinction with V extinction $A_{V}=0.5~{\rm mag}$, where they assume $A_{V}$ of the planet is the same with that of the central star AB~Aur \citep{lop06} and is a constant. The H$\alpha$ luminosity is then $L_{\rm H\alpha}=2.2\pm0.7\times10^{5}~L_{\odot}\times10^{-0.4*0.5}=1.4\pm0.4\times10^{5}~L_{\odot}$.}. Considering the typical size frequency distribution of the small grains ($dN/da\propto a^{-3.5}$), the extinction for H$\alpha$ emission of embedded planets is two times larger than that of H band (see Fig. 7 of \citet{ala24}). The H$\alpha$ extinction of AB~Aur~b is then
\begin{equation}
A_{\rm H\alpha}=2.5\log(e)\tau_{\rm H}\times2=2.82.
\label{AHalpha}
\end{equation}
The actual H$\alpha$ luminosity is then $10^{0.4A_{\rm H\alpha}}=13.4$ times larger than the 
extinction free case, $L_{\rm H\alpha}=1.9\pm0.6\times10^{-4}~L_{\odot}$. We then estimate the gas accretion rate reproducing the corrected H$\alpha$ luminosity using the relationship for planets between $L_{\rm H\alpha}$ and $L_{\rm acc}$ obtained by \citet{aoy21},
\begin{equation}
\log_{10}(L_{\rm acc}/L_{\odot})=0.95\times\log_{10}(L_{\rm H\alpha}/L_{\odot})+1.61,
\label{Lacc-LHalpha-planets}
\end{equation}
and the relationship for stars obtained by \citet{alc17},
\begin{equation}
\log_{10}(L_{\rm acc}/L_{\odot})=1.13\times\log_{10}(L_{\rm H\alpha}/L_{\odot})+1.74,
\label{Lacc-LHalpha-stars}
\end{equation}
where $L_{\rm acc}$ is the accretion luminosity (expressed as $L_{\rm shock}$ in \citetalias{shi24}).

\begin{figure}[tbp]
\centering
\includegraphics[width=\linewidth]{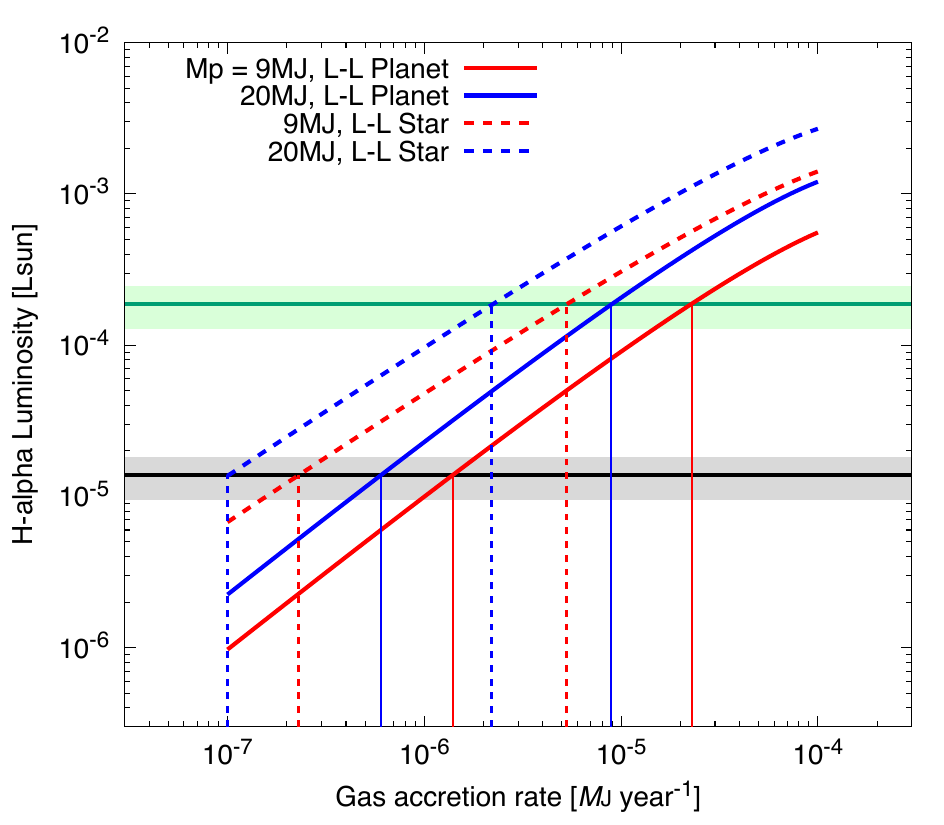}
\caption{Expected H$\alpha$ luminosity with variable gas accretion rate. The solid and dashed curves represent the cases with the $L_{\rm H\alpha}-L_{\rm acc}$ relationships for planets (Eq. (\ref{Lacc-LHalpha-planets})) and stars (Eq. (\ref{Lacc-LHalpha-stars})), respectively. The red and blue curves represent the profiles with $M_{\rm p}=9$ and $20~M_{\rm J}$, respectively. The green and black lines are the observed H$\alpha$ luminosity with $A_{\rm H\alpha}=0$ and $2.82$, respectively \citep{zho22}. The vertical solid lines represent the corresponding estimated gas accretion rates, $\dot{M}_{\rm g}=0.60$, $1.4$, $8.9$, and $23~M_{\rm J}~{\rm Myr}^{-1}$. The vertical dashed lines are $\dot{M}_{\rm g}=0.10$, $0.23$, $2.2$, and $5.3~M_{\rm J}~{\rm Myr}^{-1}$.} \label{fig:LHalpha}
\end{figure}

Figure \ref{fig:LHalpha} represents the estimated and observed H$\alpha$ luminosity. When we use the $L_{\rm H\alpha}-L_{\rm acc}$ relationships for planets (\ref{Lacc-LHalpha-planets})), the estimated gas accretion rates with the extinction are $\dot{M}_{\rm g}=2.3\times10^{-5}$ and $8.9\times10^{-6}~M_{\rm J}~{\rm yr}^{-1}$ when the planet mass is $M_{\rm p}=9$ and $20~M_{\rm J}$, respectively. These values are about 10 times larger than those without the extinction correction, $\dot{M}_{\rm g}=1.4\times10^{-6}$ and $6.0\times10^{-7}~M_{\rm J}~{\rm yr}^{-1}$, respectively. We note that the estimated gas accretion rate with no extinction correction and $M_{\rm p}=9~M_{\rm J}$ is consistent with the value obtained in \citetalias{Currie2022}. On the other hand, when we use the $L_{\rm H\alpha}-L_{\rm acc}$ relationships for stars (Eq. (\ref{Lacc-LHalpha-stars})), the gas accretion rates with the extinction are estimated as $\dot{M}_{\rm g}=5.3\times10^{-6}$ and $2.2\times10^{-6}~M_{\rm J}~{\rm yr}^{-1}$ when the planet mass is $M_{\rm p}=9$ and $20~M_{\rm J}$, respectively. Without the extinction, the corresponding estimated gas accretion rates are $\dot{M}_{\rm g}=2.3\times10^{-7}$ and $1.0\times10^{-7}~M_{\rm J}~{\rm yr}^{-1}$, which are about 10 times smaller than those considering the extinctions.

\begin{figure*}[tbp]
\centering
\includegraphics[width=0.49\linewidth]{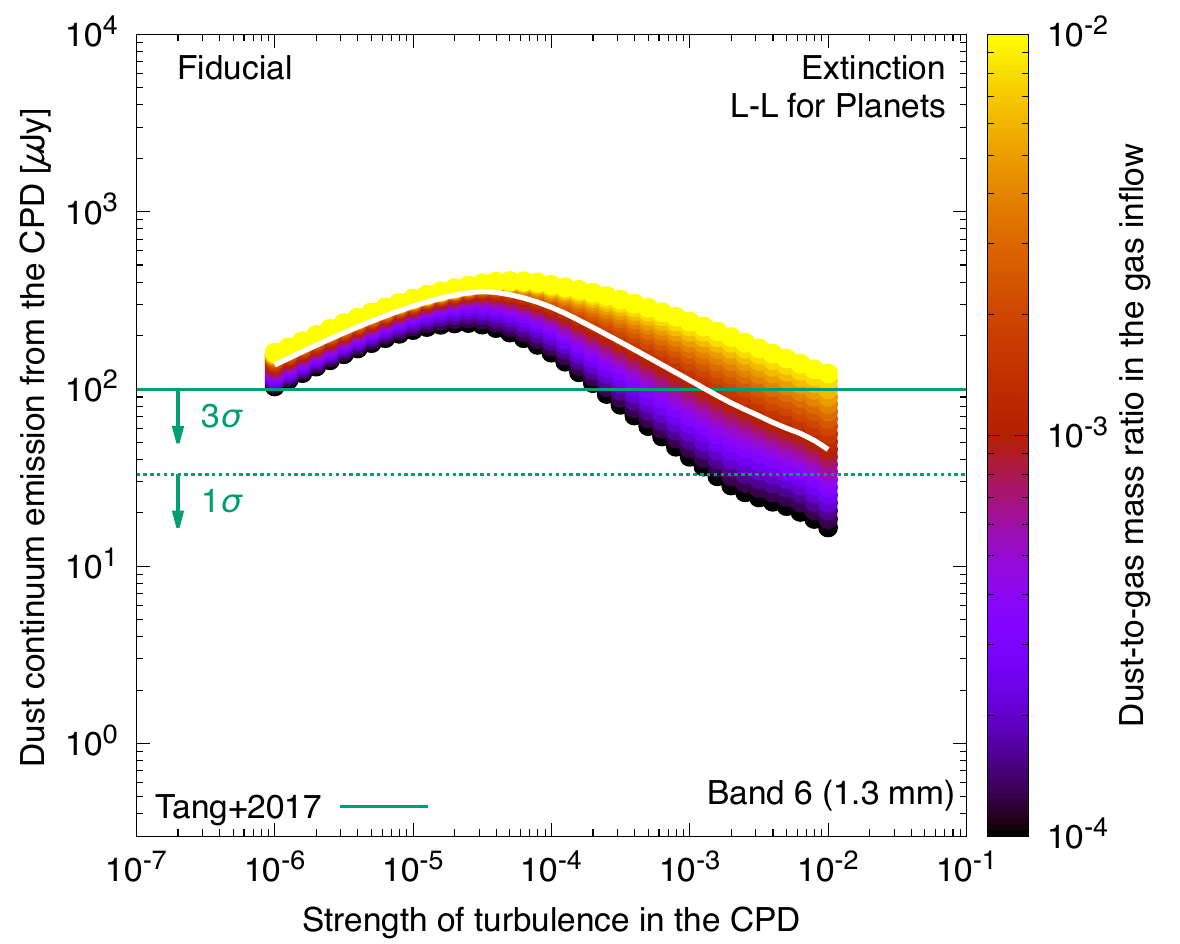}
\includegraphics[width=0.49\linewidth]{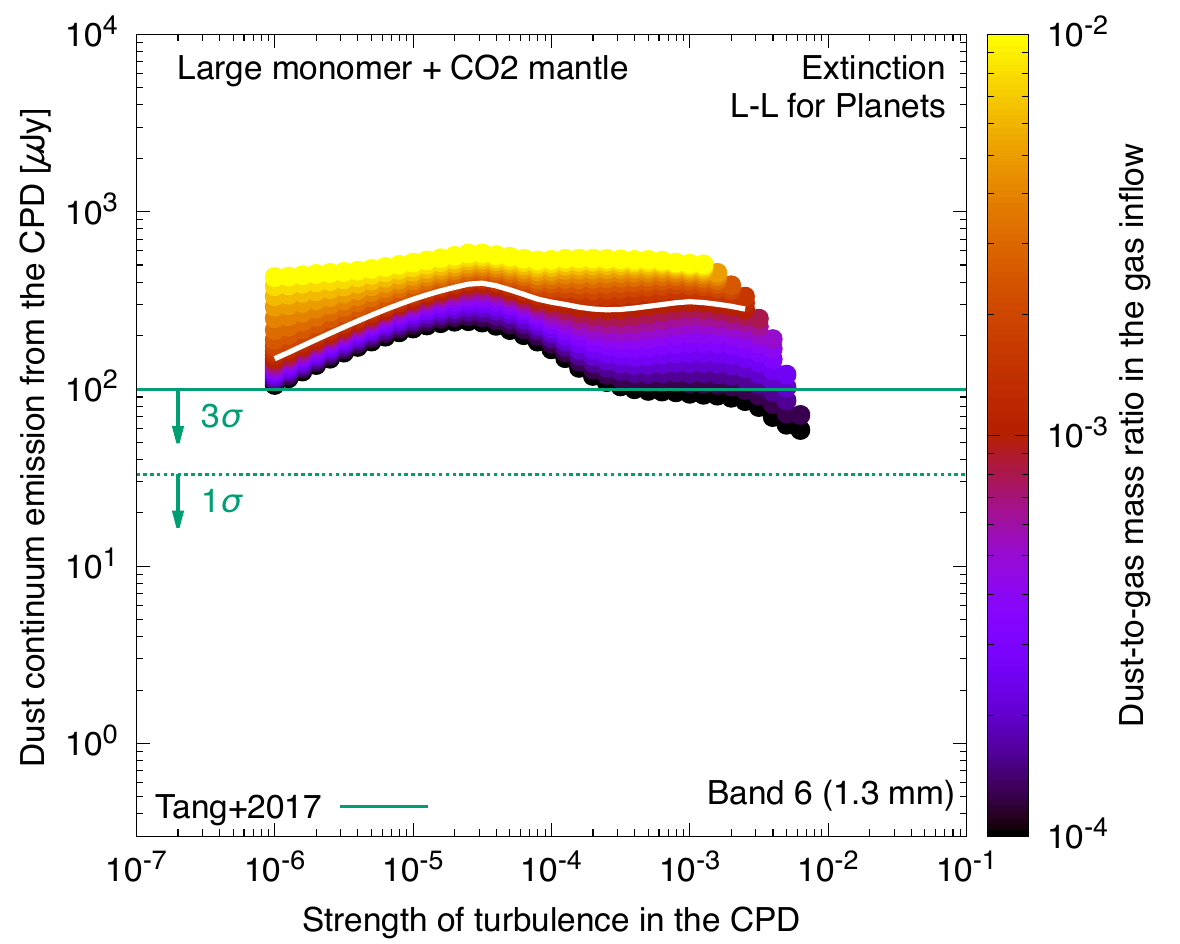}
\includegraphics[width=0.49\linewidth]{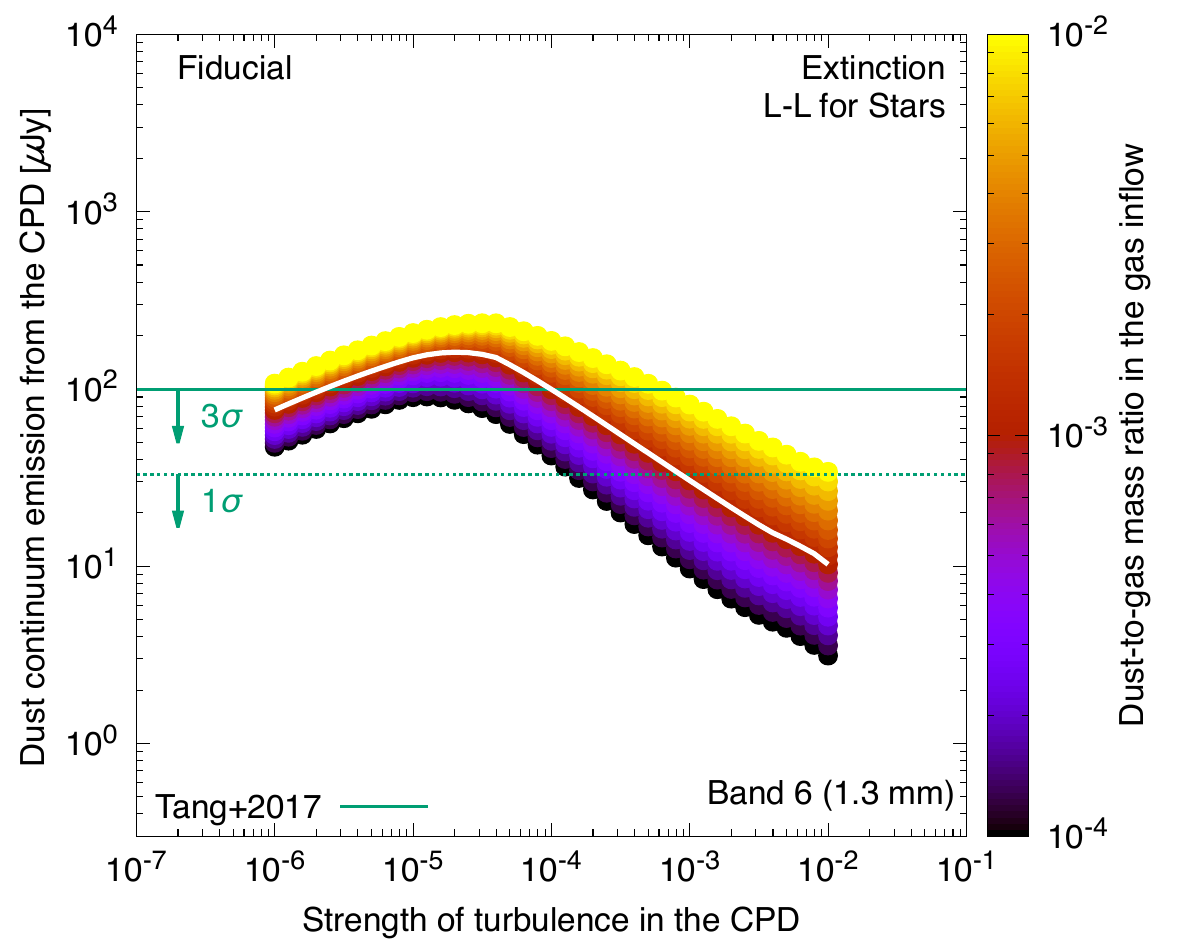}
\includegraphics[width=0.49\linewidth]{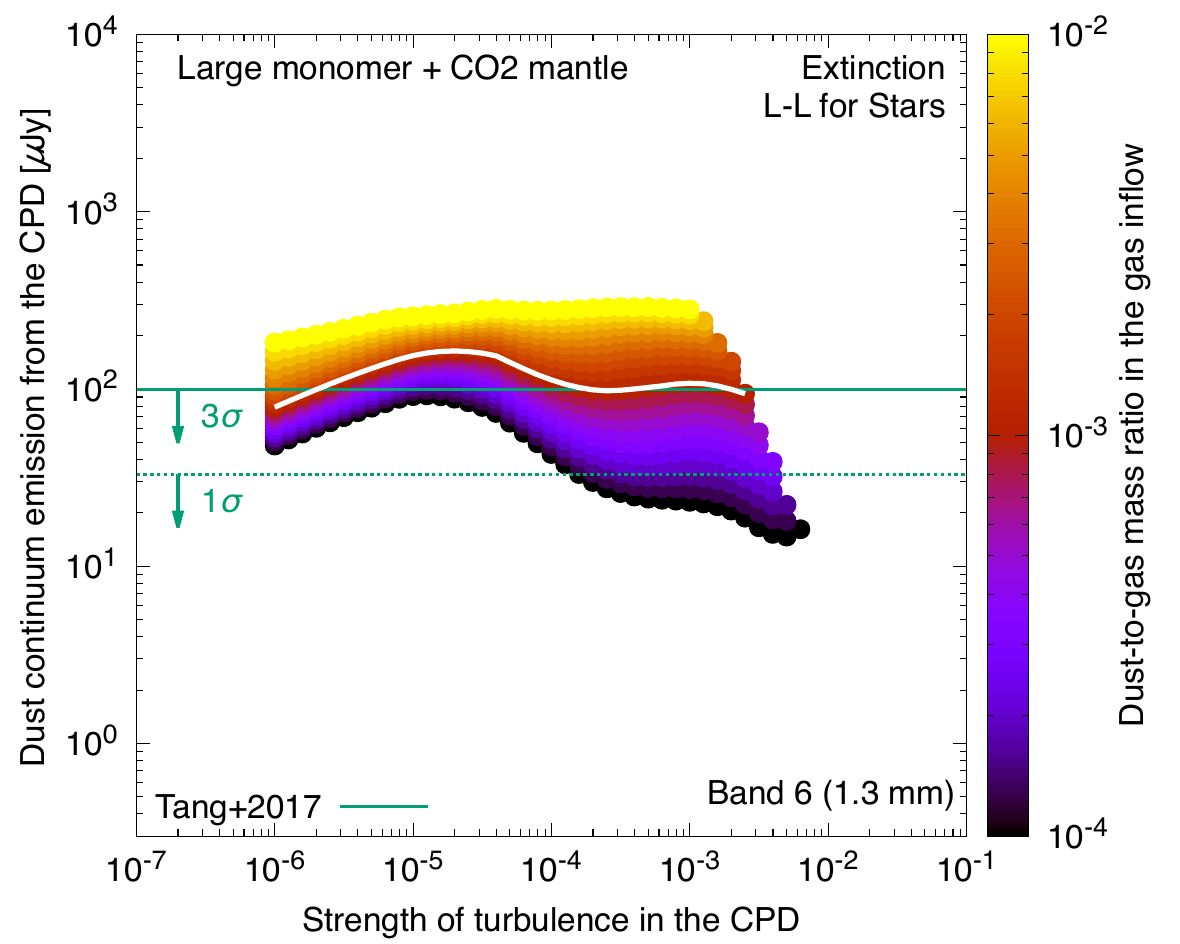}
\caption{Same as Fig. \ref{fig:prediction} but considering the reduction of the observed emission due to the extinction by small grains (\texttt{Extinction}). The left and right panels represent the \texttt{Fiducial} and \texttt{Large monomer + CO$_{2}$ mantle} cases. The upper panels are the cases using the $L_{\rm H\alpha}-L_{\rm acc}$ relationships for planets (Eq. (\ref{Lacc-LHalpha-planets})); $M_{\rm p}=20~M_{\rm J}$ and $\dot{M}_{\rm g}=8.9\times10^{-6}~M_{\rm J}~{\rm yr}^{-1}$. The lower panels are the cases using the $L_{\rm H\alpha}-L_{\rm acc}$ relationships for stars (\ref{Lacc-LHalpha-stars})); $M_{\rm p}=20~M_{\rm J}$ and $\dot{M}_{\rm g}=2.2\times10^{-6}~M_{\rm J}~{\rm yr}^{-1}$. We note that some results with $\alpha\sqrt{x}\gtrsim10^{-4}$ are not presented in the right panels because of numerical reasons.
} \label{fig:extinction}
\end{figure*}

We then estimate the expected flux density of the dust emission from the CPD when the planet mass and the gas accretion rate are the values obtained from the observed emission at H band and the H$\alpha$ line emission considering the extinction by small grains; $M_{\rm p}=20~M_{\rm J}$ and $\dot{M}_{\rm g}=8.9\times10^{-6}$ or $2.2\times10^{-6}~M_{\rm J}~{\rm yr}^{-1}$ (\texttt{Extinction})\footnote{We note that even if we chose the $9~M_{\rm J}$ cases, $F_{{\rm d},\lambda}$ should not be different from that of the $20~M_{\rm J}$ cases, because both $F_{{\rm d},\lambda}$ and $L_{\rm H\alpha}$ are roughly proportional to $M_{\rm p}\dot{M}_{\rm g}$ (see Eqs. (\ref{Lacc-LHalpha-planets}), (\ref{Lacc-LHalpha-stars}), and \citetalias{shi24}).}. The upper panels of Fig. \ref{fig:extinction} shows that the expected dust emission of \texttt{Extinction} with the $L_{\rm H\alpha}-L_{\rm acc}$ relationship for planets is about 10 times larger than that without the extinction correction and with any conditions of the monomers. With the $L_{\rm H\alpha}-L_{\rm acc}$ relationship for stars, $F_{{\rm d},\lambda}$ is about four times larger. These results are because the flux density from the CPD is about proportional to the gas accretion rate, which is consistent with the cases of PDS~70~b and c (\citetalias{shi24}). As a result, with the \texttt{Fiducial} cases (left panels), the dust emission with the typical dust-to-gas mass inflow ratio, $x=0.001$, is higher than the $3~\sigma$ of the previous observation \citet{Tang2017} when $\alpha\lesssim2\times10^{-3}$ and $\alpha\lesssim10^{-4}$ for the cases using the $L_{\rm H\alpha}-L_{\rm acc}$ relationships for planet and stars, respectively. In the cases of \texttt{Large monomer + CO$_{2}$} (right panels), the expected dust emission with $x=0.001$ is higher than the $3\sigma$ with (at least) $\alpha\lesssim2\times10^{-3}$ when both the $L_{\rm H\alpha}-L_{\rm acc}$ relationships for planets and stars are used. In other words, if the effects of small grains are considered, the non-detection of the dust emission from AB~Aur~b is only explained when $x<0.001$ in a broad range of $\alpha$ value.

However, we note that the $20~M_{\rm J}$ object (which should be classified as a brown dwarf) would perturb the circumstellar material of AB~Aur beyond what has been revealed in previous observations. One possible reason why the $20~M_{\rm J}$ object has not caused a significant gap might be that it has only recently formed. It takes time to perturb the disk and open a deep gap. On the other hand, spirals arms can be excited on dynamical timescale, and the spirals inside AB Aur b’s location might not be as prominent as we may expect from a $20~M_{\rm J}$ object.

We also note that, even if we assume that $M_{\rm p}=9~~M_{\rm J}$ and 
$\dot{M}_{\rm g}=2.3\times10^{-5}~M_{\rm J}~{\rm yr}^{-1}$ or $5.3\times10^{-6}~M_{\rm J}~{\rm yr}^{-1}$ (i.e., all contribution of the increase of $M_{\rm p}\dot{M}_{\rm g}$ due to the extinction is distributed to $\dot{M}_{\rm g}$ and nothing to $M_{\rm p}$), the estimated flux density of the dust emission from the CPD should be almost the same with the results shown in Fig. \ref{fig:extinction}, because the flux density is almost proportional to $M_{\rm p}\dot{M}_{\rm g}$.

\subsection{Why is the potential CPD not detected?} \label{sec:planet}
There have not been any detection of dust continuum at AB~Aur~b by ALMA observations, and we found that the expected flux density of the dust emission from the potential CPD of AB~Aur~b is higher than the three r.m.s. noise level of the previous observation when the upward revision of the planet mass and the gas accretion rate caused by the reduction of the observed near-infrared continuum and H$\alpha$ line emission due to the extinction by micron sized grains is considered. There are three possibilities\footnote{We note that the signal-to-noise ratio of the observation by \citet{Tang2017} can be low depending on the point spread function, which is a potential reason of the non-detection as well.}.

The first possibility is that there is a CPD, but the dust supply is very limited. Fig. \ref{fig:extinction} shows that if the dust-to-gas mass ratio in the inflow is $x<0.001$, the expected dust flux density should be lower than the $3\sigma$ even if the extinction by small grains are considered. This extremely low $x$ value can be achieved when the gap around the planet is deep and its outer edge's pressure gradient is steep enough, or the vertical and horizontal diffusion of dust at the gap edge is extremely small, or the vertical diffusion of dust inside the gap (i.e., the root of the gas inflow) is small \citep{hom20}. However, this situation is not qualitatively consistent with the assumption of the \texttt{Extinction} case; there is an enough amount of small grains (i.e., dust) to cause the extinction. Therefore, if the presence of the planet and its CPD are accepted, the amount of dust supply to the vicinity of the planet must be small, which does not cause the dust extinction at micrometer wavelengths and has weak dust emission at (sub)millimeter wavelength.

The second possibility is that there is not a CPD but an envelope. Gas around a planet first forms an envelope-like structure during the gas accretion of the planet. Once the gas envelope cools enough, the mechanism to support the gas against the gravitational collapse shifts from the heat pressure to the centrifugal force, resulting in the formation of a CPD around the planet \citep[e.g.,][]{ayl09a,szu16}. According the previous hydrodynamical simulations including radiative transfer models, the opacity plays an important role, which is determined by the amount of dust supplied to the vicinity of the planet. As the amount of dust is larger, the opacity increases, and the cooling timescale of the gas becomes longer, which makes the formation of the CPD more difficult \citep{fun19,kra24}. Dust particles in an envelope (maybe not grown large) also emit thermal emission depending on the temperature, but the emission at (sub)millimeter wavelength is likely to be weaker than that of a CPD in robust situations \citep{cho24}. Therefore, even if the amount of dust supply to the vicinity of the planet is large, the non-detection may be possible if the planet does not have a CPD but an envelope. At the same time, the emission is weak in both envelope and CPD cases as the dust supply is small. Since the flux density of the dust emission from an envelope cannot be estimated by our CPD and dust evolution models, detailed comparisons between the envelope and CPD cases are the future works.

The third possibility is that there is no gas accreting planet but just scattered light of the central star. Since polarized intensity imaging did not show a concentration at the location, \citetalias{Currie2022} has argued that AB~Aur~b is a gas accreting planet, but it is still controversial. \citet{zho22} has claimed that the detected H$\alpha$ signal, about three times higher than the continuum brightness, may have identified a scattered light component. H$\alpha$ line emission in excess of continuum emission is usually considered as an evidence of a gas accreting planet; it is interpreted that gas falls onto a (proto)planet (or a CPD) and makes gas shocks producing high temperature environments where H$\alpha$ can emits. However, in the case of AB~Aur~b, the central star AB~Aur itself has strong H$\alpha$ emission, about 2.4 times higher than the continuum. Moreover, \citet{zho23} shows that their ultraviolet (UV) and optical observations by the HST Wide Field Camera 3 do not necessitate the presence of a planet. \citet{bid24} does not detect significant Pa$\beta$ emission from AB~Aur~b, where such a higher-order emission line should also be present. They argue that the non-detection suggests a low gas accretion rate, but we also note that their estimate may be underestimated due to an inaccurate source model of AB~Aur~b \citep{cur24}.

The non-detection of the dust emission from the potential CPD is consistent with the idea that AB~Aur~b is scattered light, but it is difficult to exclude the possibility of a gas accreting planet. There are still some ways to explain the non-detection with the existence of the planet as we discussed above in this section.

\subsection{Expected dust emission at ALMA Band 7} \label{sec:band7}
\begin{figure*}[tbp]
\centering
\includegraphics[width=0.49\linewidth]{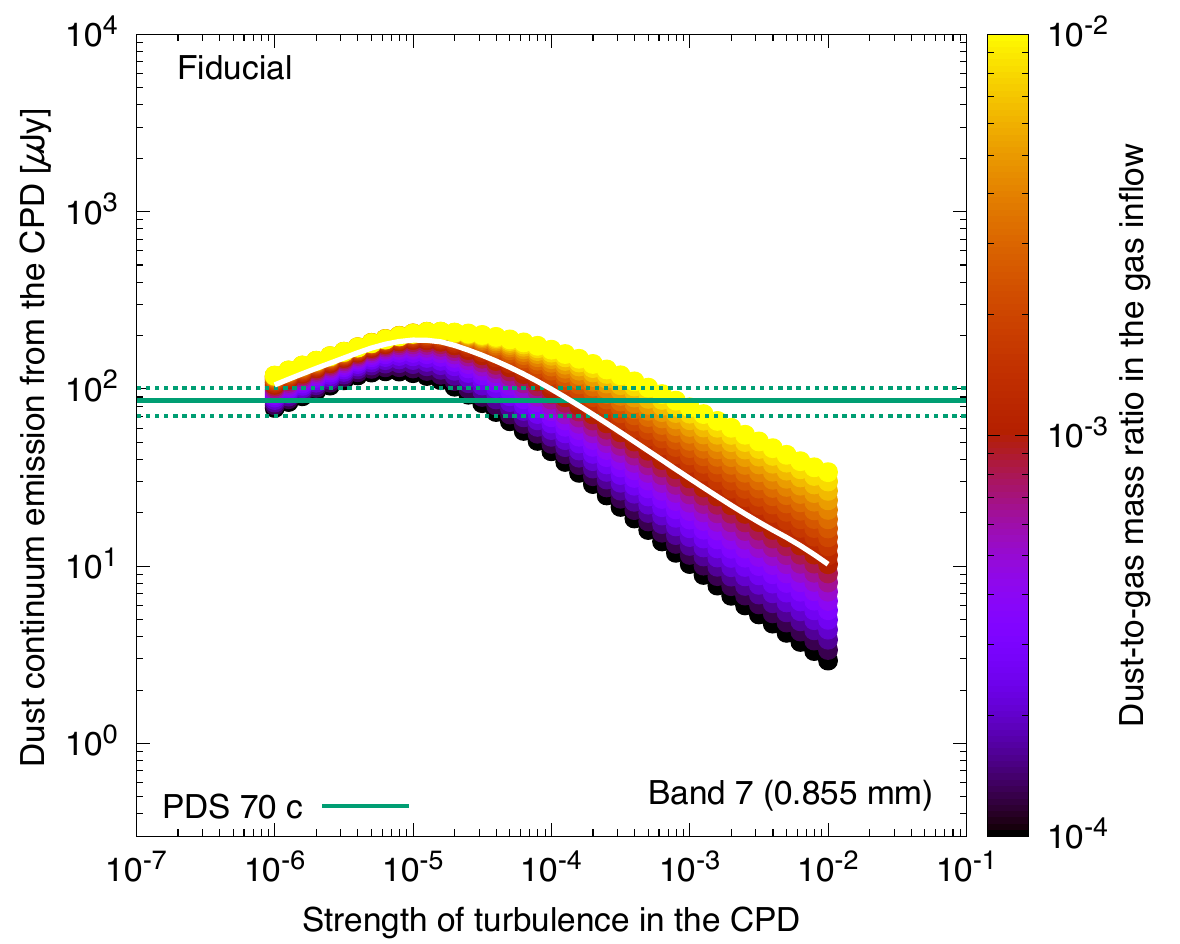}
\includegraphics[width=0.49\linewidth]{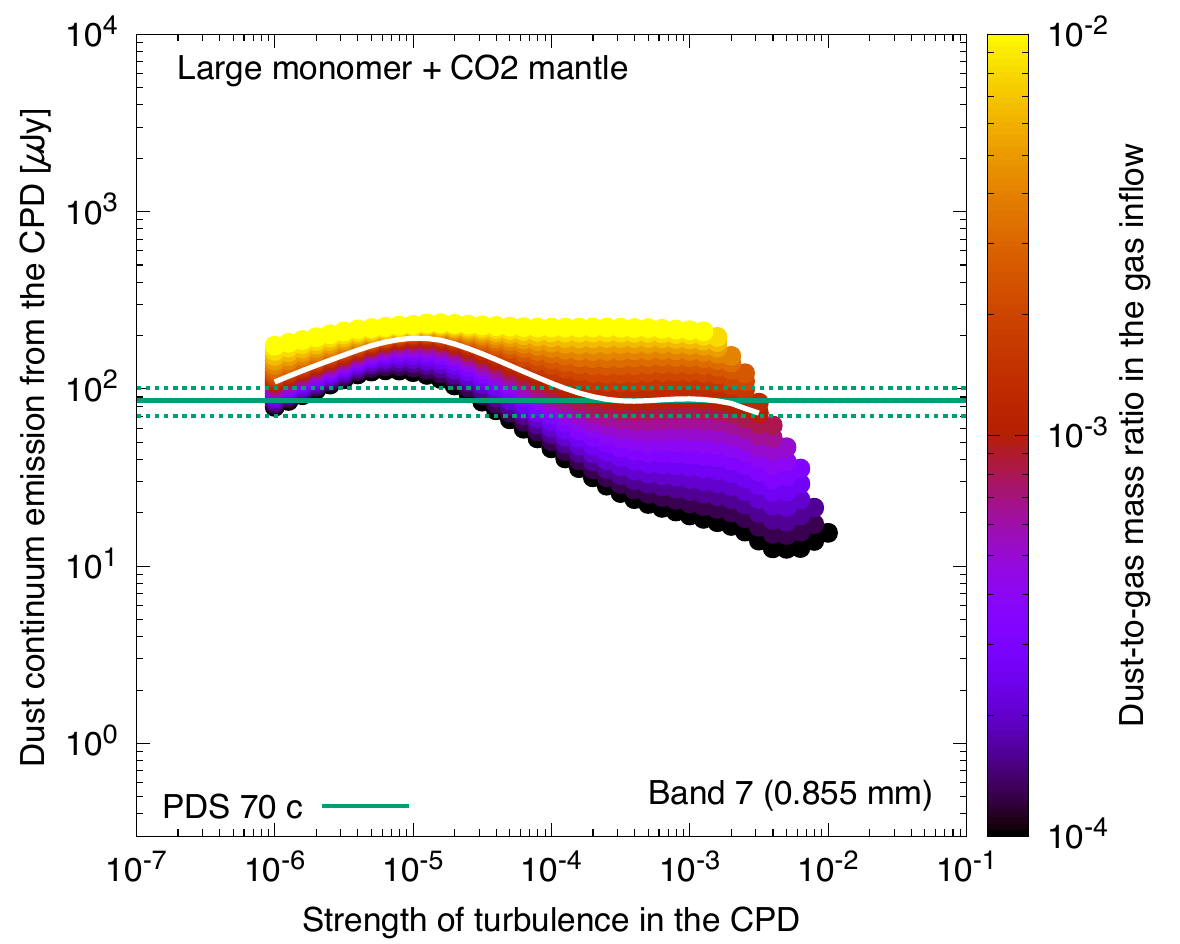}
\includegraphics[width=0.49\linewidth]{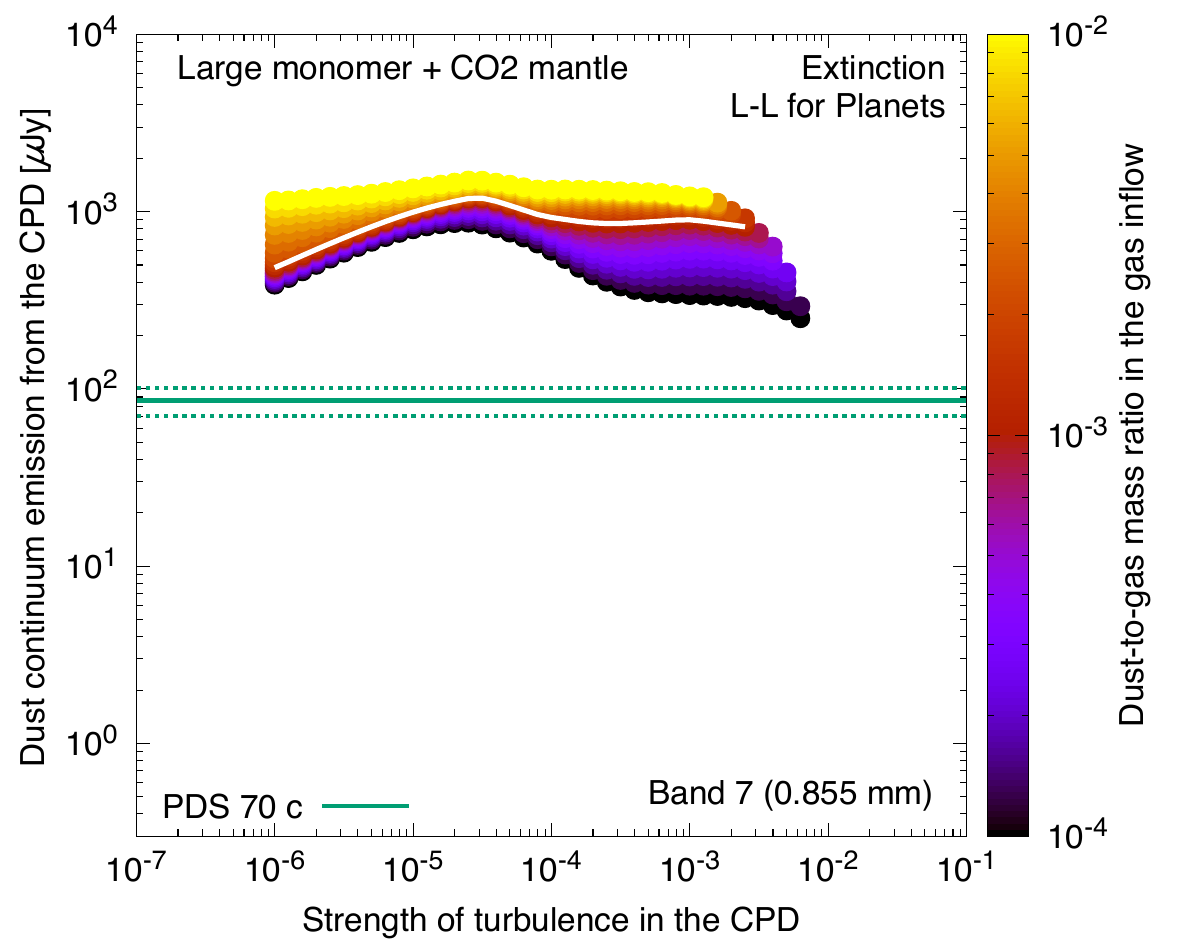}
\includegraphics[width=0.49\linewidth]{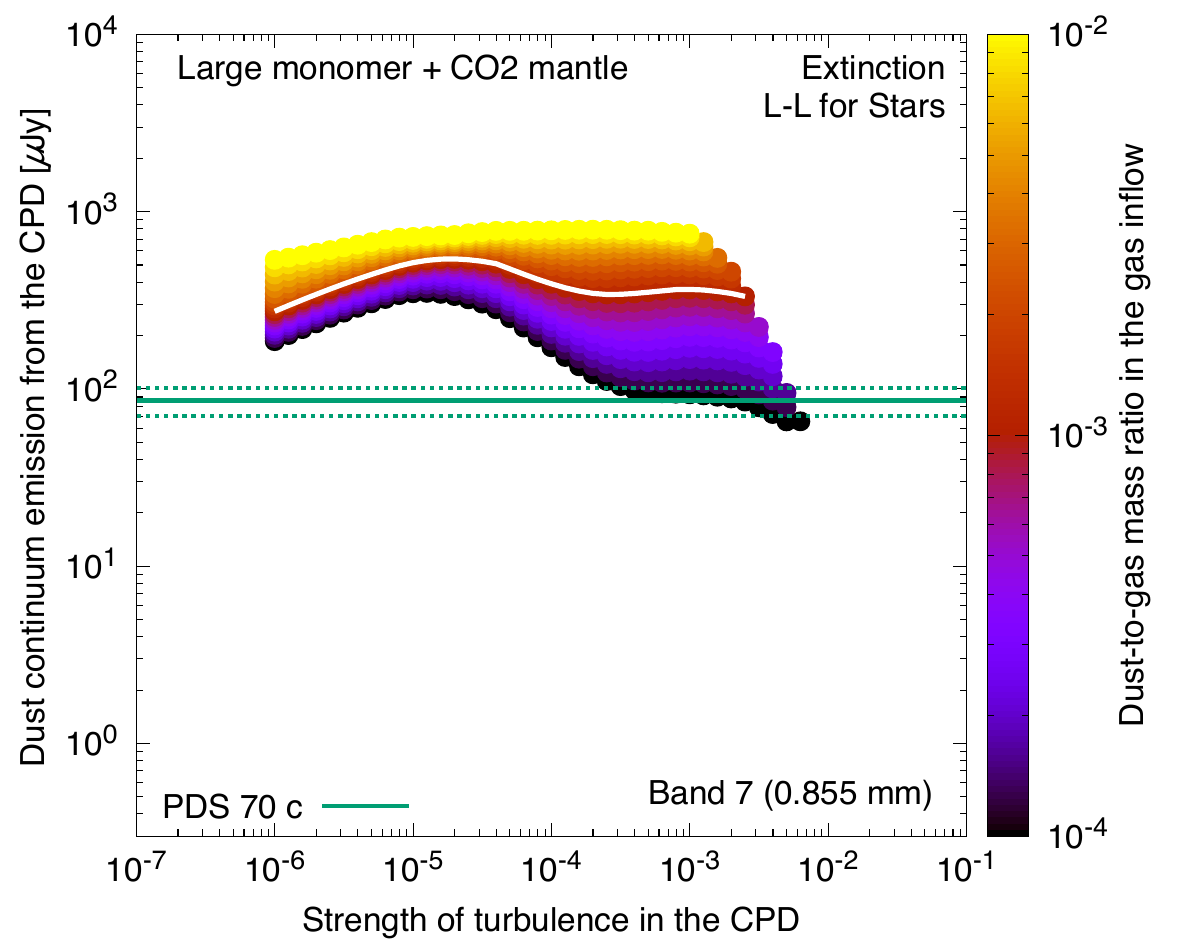}
\caption{Same as the \texttt{Fiducial} and \texttt{Large monomer + CO$_{2}$ mantles} cases of Figs. \ref{fig:prediction} and \ref{fig:extinction} but at Band 7 ($\lambda=855~\mu{\rm m}$). The left and right lower panels represent the \texttt{Extinction} cases with \texttt{Large monomer + CO$_{2}$ mantles} using the $L_{\rm H\alpha}-L_{\rm acc}$ relationships for planets (Eq. (\ref{Lacc-LHalpha-planets})) and for stars (\ref{Lacc-LHalpha-stars})), respectively. The green lines represent the observed value of PDS~70~c, $86\pm16~\mu{\rm Jy}$ \citep{Benisty2021}. We note that some results with $\alpha\sqrt{x}\gtrsim10^{-4}$ are not presented in the upper right, lower, lower left, and lower right panels because of numerical reasons.
} \label{fig:band7}
\end{figure*}

Is it possible to investigate the probability of the presence of an accreting planet further from the CPD observation aspect? One way is to observe the dust continuum from the planet candidate and its possible CPD by other (sub)millimeter wavelengths. Since the Planck function has negative dependence on the wavelength around ALMA's observation wavelengths, the expected dust continuum emission is larger when the wavelength is shorter than Band 6 (see Eq. (\ref{Fdlambda})). Figure \ref{fig:band7} shows the expected flux density of the dust continuum emission at Band 7 ($\lambda=855~\mu{\rm m}$). The upper left panels shows that the expected value is consistent with or higher than the detected flux density of the dust emission from the CPD of PDS~70~c, $86\pm16~\mu{\rm Jy}$ \citep{Benisty2021} when $\alpha\lesssim10^{-3}$ and $x=0.001$. When the monomers of dust are large and covered by CO$_{2}$ mantles (\texttt{Large monomer + CO$_{2}$ mantle}), the expected emission is consistent with the observed value of PDS~70~c (at least) when $\alpha\lesssim2\times10^{-3}$ (upper right). Moreover, if the reduction of the observed near-infrared and H$\alpha$ emission due to the extinction by small grains is considered (\texttt{Extinction}), the expected dust emission is higher than the detected value of PDS~70~c with (at least) $\alpha\lesssim3\times10^{-3}$ and with both the $L_{\rm H\alpha}-L_{\rm acc}$ relationships for planet and stars. Therefore, future continuum observations with ALMA at Band 7 are preferable to detect the dust emission from the potential CPD of AB~Aur~b or to obtain a strong suggestion to the presence of a gas accreting planet.

\section{Conclusions} \label{sec:conclusions}
Thermal emission from the dust in the CPD of a gas accreting planet PDS~70~c has been detected at ALMA Band 7 ($\lambda=855~\mu{\rm m}$) \citep{isell19a,Benisty2021}. However, dust emission from AB~Aur~b, recently reported as another gas accreting planet, has not been detected at ALMA Band 6 ($1.3~{\rm mm}$) \citep{Tang2017}, and there is still discussion whether AB~Aur~b is a planet or scattered light of the central star at the surface of the PPD \citep{zho22,zho23,bid24,cur24}. We calculated the evolution of dust in the potential CPD of AB~Aur~b and predicted the flux density of its thermal emission as a case study by updating the model produced by \citetalias{shi24}. We showed that the expected flux density is lower than the $3~\sigma$ noise level of the previous observation, $99~\mu{\rm Jy}$, when we assume the planet properties as the values estimated by the observations of SED of near-infrared in \citetalias{Currie2022}. We also found that the dependence of the predicted flux density of the dust emission on the strength of turbulence in the CPD is weak when the monomers of the dust are large ($a_{\rm mon}=1.5~\mu{\rm m}$) and covered with CO$_{2}$ mantles, because they are more fragile compared to normal icy particles. This feature may also be applied to the potential CPDs of other gas accreting planets.

However, \citetalias{Currie2022} also predicted that the planet candidate is embedded in small grains. We considered the effects of the reduction of the observed near-infrared continuum and H$\alpha$ line emission due to the extinction by the small grains; the planet mass and the gas accretion rate can be much larger than the previously estimated values in \citetalias{Currie2022}. We recalculated the flux density of dust emission with these corrected values and found that the flux density is higher than the $3~\sigma$ of the previous observation with a broad range of the strength of turbulence in the CPD and with the typical dust-to-gas mass ratio in the inflow to the CPD, $x=0.001$. This result suggests that the amount of the supply of dust to the vicinity of the planet candidate is small if the planet has a CPD. We also predicted the flux density of dust continuum at ALMA Band 7 and showed that it would be much stronger than the detected dust emission from the CPD of PDS~70~c \citep{Benisty2021}. Therefore, future continuum observations at ALMA Band 7 are preferable to obtain more robust clues to the question whether AB~Aur~b is a planet or not.

\acknowledgments
We thank the anonymous referee for the very useful comments improving our manuscript a lot. We also thank Munetake Momose for helpful and constructive discussion. This work was supported by JSPS KAKENHI Grant Numbers JP19H00703, JP19H05089, JP19K03932, JP22H01274, JP23K22545 and JP24K22907. This work has been carried out within the framework of the NCCR PlanetS supported by the Swiss National Science Foundation under grants 51NF40\_182901 and 51NF40\_205606. C.M. acknowledges the funding from the Swiss National Science Foundation under grant 200021\_204847 `Planets In Time'.

\bibliography{ABAurb-dust}
\bibliographystyle{aasjournal}

\appendix
\section{Previous continuum observation of AB~Aur~b} \label{sec:previous}
\begin{figure}[ht]
\centering
\includegraphics[width=\linewidth]{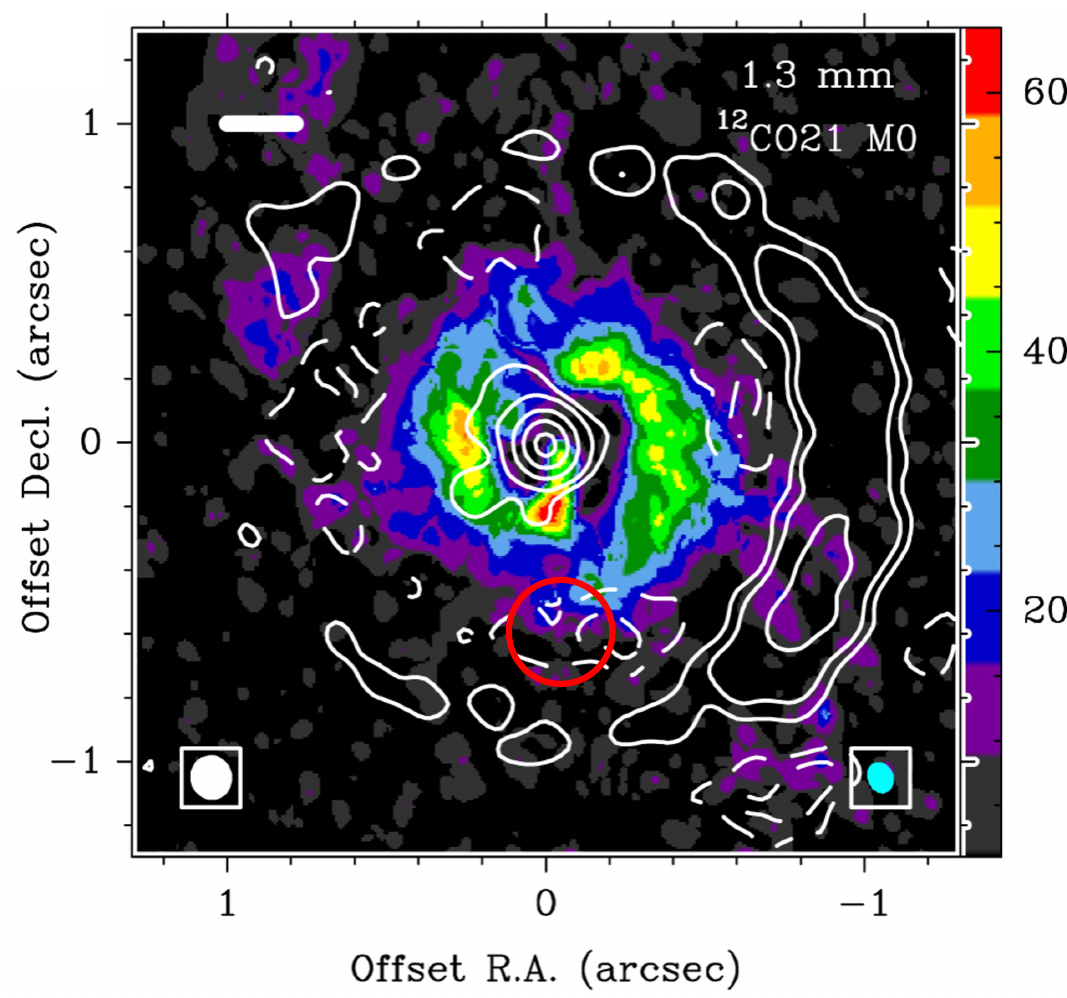}
\caption{Band 6 (1.3 mm) continuum emission (contour) and the moment 0 map of CO (color scale; units of mJy/beam km $s^{-1}$, where the beam is $0.11\arcsec$ (sky blue ellipse at the right bottom corner)). The contours are $-6\sigma$, $-3\sigma$, $3\sigma$, $15\sigma$, $30\sigma$, and $50\sigma$, where $1\sigma$ is $33~\mu{\rm Jy}$ (0.04 K) per $0.14\arcsec$ beam (white ellipse at the left bottom corner). This figure is cited from the panel (b) of Figure 1 in \citet{Tang2017}. We added a red circle denoting the location of AB~Aur~b \citep{Currie2022}.}
\label{fig:ABAurb}
\end{figure}

Figure \ref{fig:ABAurb} shows the Band 6 (1.3 mm) continuum emission (contour) and the moment 0 map of CO (color scale), cited from \citet{Tang2017}. Inside the red circle, the location of AB~Aur~b \citep{Currie2022}, there is no detection of dust continuum emission. The noise level of the observation is $33~\mu{\rm Jy}$ per $0.14\arcsec$ beam. The beam size is clearly larger than the size of the  dust-containing region of the CPD of AB~Aur~b ($\sim0.01\arcsec$) and smaller than the distance between AB~Aur~b and the outer dust ring. Since the flux density of the dust emission at the location of AB~Aur~b is negative in Fig. \ref{fig:ABAurb}, we interpret this as indicating that the dust emission from the potential CPD is zero (i.e., non-detection), and so we compare our model predictions with $\sigma=33~\mu{\rm Jy}$ and $3\sigma=99~\mu{\rm Jy}$.

\section{Approximation of the dust-to-gas surface density ratio for the disk temperature calculations} \label{sec:opacity}
\begin{figure}[ht]
\centering
\includegraphics[width=\linewidth]{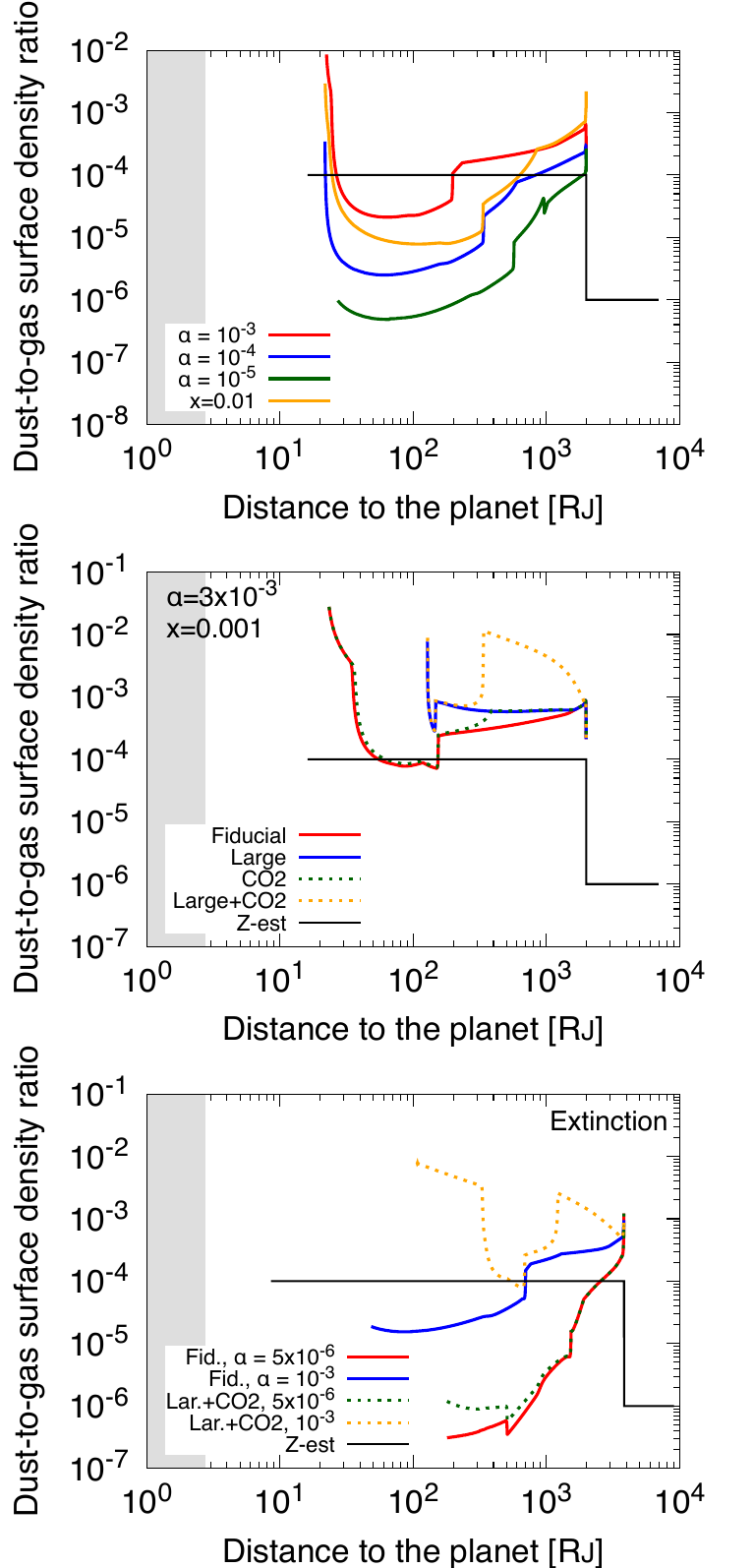}
\caption{Radial distribution of $Z_{\rm\Sigma}$ (solid color curves) and $Z_{\rm\Sigma,est}$ (black lines) with the various parameter sets. The top, middle, bottom panels show the profiles with the \texttt{Fiducial} cases, the cases considering the effects of the monomer conditions, and the \texttt{Extinction} cases with the $L_{\rm H\alpha}-L_{\rm acc}$ relationship for planets ($M_{\rm p}=20~M_{\rm J}$ and $\dot{M}_{\rm g}=8.9\times10^{-6}~M_{\rm J}~{\rm yr}^{-1}$), respectively. The colors of the curves in the top and middle panels are consistent with those in Figs. \ref{fig:dust-distribution} and \ref{fig:conditions}, respectively. The bottom panel shows the profiles of \texttt{Extinction} with the $L_{\rm H\alpha}-L_{\rm acc}$ relationship for planets. The red and blue profiles represent the \texttt{Fiducial} cases with $\alpha=5\times10^{-6}$ and $10^{-3}$, respectively. The green and orange profiles are the corresponding \texttt{Large monomer + CO$_{2}$ mantle} cases.}
\label{fig:Zest}
\end{figure}

\begin{figure}[ht]
\centering
\includegraphics[width=0.98\linewidth]{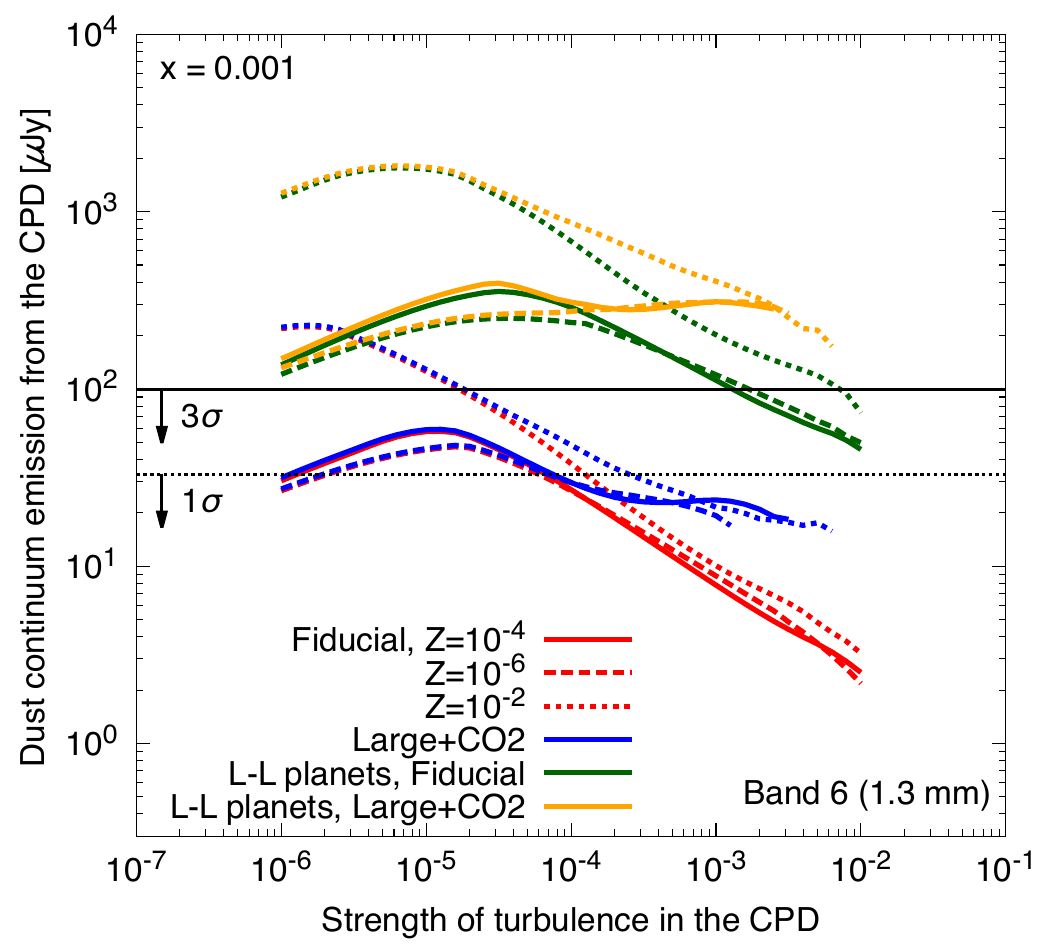}
\caption{Expected flux density of the dust continuum emission from the CPD of AB~Aur~b at ALMA Band 6 ($\lambda=1.3~{\rm mm}$) with $x=0.001$. The solid, dashed, and dotted curves represent the cases with $Z_{\Sigma,{\rm est}}=10^{-4}$, $10^{-6}$, and $10^{-2}$, respectively. The red and blue curves are the \texttt{Fiducial} and \texttt{Large monomer + CO$_{2}$} cases without the extinction effects, respectively. The green and orange curves are  the \texttt{Fiducial} and \texttt{Large monomer + CO$_{2}$} cases with  \texttt{Extinction} of the $L_{\rm H\alpha}-L_{\rm acc}$ relationship for planets, respectively. The black solid and dotted horizontal lines represent the noise levels of $3~\sigma=99~\mu{\rm Jy}$ and $1~\sigma~\mu{\rm Jy}=33~\mu{\rm Jy}$ of the previous observation by \citet{Tang2017} (see Appendix \ref{sec:previous})}, respectively.
\label{fig:Zest-depencdence}
\end{figure}

In this work, the radial distribution of the surface density and size of dust is calculated, but the opacity used for the calculations of the disk temperature is not calculated simultaneously. We fix the dust-to-gas surface density ratio used for the calculation of the opacity in the gas disk model as $Z_{\Sigma,{\rm est}}=10^{-4}$ in the dust-containing region (i.e., gas inflow region) of the CPD, $r\leq r_{\rm inf}$, and $Z_{\Sigma,{\rm est}}=10^{-6}$ at $r>r_{\rm inf}$. Figure \ref{fig:Zest} shows that this assumed value is almost consistent or lower than the subsequently calculated value, $Z_{\Sigma}=\Sigma_{\rm d}/\Sigma_{\rm g}$, in the dust evolution model. On the other hand, Figure \ref{fig:Zest-depencdence} shows that the $Z_{\Sigma,{\rm est}}$ dependence of the flux density of the dust emission $F_{{\rm d},\lambda}$ is weak; the difference of two orders of magnitude in $Z_{\Sigma,{\rm est}}$ only makes difference of $F_{{\rm d},\lambda}$ by a factor of two with most $\alpha$. Therefore, our simple estimate is reasonable. When $\alpha$ is small, $F_{{\rm d},\lambda}$ is larger as $Z_{\Sigma,{\rm est}}$ is large. However, this difference should not change our conclusion of this work that the non-detection suggests small dust supply to the vicinity of AB~Aur~b either, because we assume a smaller value of $Z_{\Sigma,{\rm est}}$ than the calculated value of $Z_{\Sigma}$ (Fig. \ref{fig:Zest}).

\end{document}